\documentclass[11pt,a4paper]{article}
\pdfoutput=1

\usepackage{jheppub}

\usepackage[utf8]{inputenc}

\usepackage{hyperref}

\usepackage{amsmath}
\usepackage{amssymb}
\usepackage{xcolor}
\usepackage{physics}
\usepackage{graphicx} 

\usepackage{pgf} 

\usepackage{tabularx}
\usepackage{multirow}
\usepackage{multicol}

\usepackage{soul}

\numberwithin{equation}{section}

\newenvironment{eqaed}
    {\begin{equation}
    \begin{aligned}
    }
    { 
    \end{aligned}
    \end{equation}
    }

\newcommand{\commie}[1]{}

\makeatletter
\gdef\@fpheader{}
\makeatother

\begin{document}

\author[a]{Ivano Basile}
\author[b]{Alessia Platania} 

\affiliation[a]{Service de Physique de l'Univers, Champs et Gravitation, Universit\'{e} de Mons, Place du Parc 20, 7000 Mons, Belgium}

\affiliation[b]{Perimeter Institute for Theoretical Physics, 31 Caroline St.  N., Waterloo, ON N2L 2Y5, Canada}
  
\emailAdd{ivano.basile@umons.ac.be}
\emailAdd{aplatania@perimeterinstitute.ca}

\title{Cosmological $\alpha'$-corrections from the functional renormalization group}

\begin{abstract}
{We employ the techniques of the Functional Renormalization Group in string theory, in order to derive an effective mini-superspace action for cosmological backgrounds to all orders in the string scale $\alpha'$. 
To this end, T-duality plays a crucial role, classifying all perturbative curvature corrections in terms of a single function of the Hubble parameter. The resulting renormalization group equations admit an exact, albeit non-analytic, solution in any spacetime dimension $D$, which is however incompatible with Einstein gravity at low energies. Within an $\epsilon$-expansion about~$D=2$, we also find an analytic solution which exhibits a non-Gaussian ultraviolet fixed point with positive Newton coupling, as well as an acceptable low-energy limit. Yet, within polynomial truncations of the full theory space, we find no evidence for an analog of this solution in~$D=4$. Finally, we comment on potential cosmological implications of our findings.}
\end{abstract}

\maketitle

\section{Introduction}\label{sec:introduction}
    
Despite the tremendous success of Einstein gravity in describing low-energy phenomena, its perturbative non-renormalizability, along with other deeply rooted subtleties, has long remained a remarkably puzzling challenge to a more complete understanding of the high-energy behavior of the theory and to the formulation of a consistent ultraviolet (UV) completion. While a number of proposals have been put forth, any candidate theory is to determine infinitely many curvature corrections to the low-energy effective action in terms of finitely many parameters.

Within the framework of asymptotically safe gravity, this would be achieved by a suitable fixed point of the renormalization group (RG) flow, whose critical surface is finite-dimensional~\cite{1976W}. One of the possible approaches to investigate this proposal is the study of the RG flow of the so-called effective average action (EAA)~\cite{Wetterich:1992yh}, a scale-dependent action functional containing all possible operators compatible with the symmetries and field content of the theory. The RG flow of the EAA is typically studied using Functional Renormalization Group (FRG) techniques~\cite{eaa,Dupuis:2020fhh}, which involve a functional integro-differential equation for the EAA~\cite{Wetterich:1992yh}. Although in principle these methods can capture the complete physics in a non-perturbative fashion, solving the FRG equation exactly is currently out of reach and the full theory space is typically restricted via truncated derivative or vertex expansions of the EAA~\cite{eaa}. These methods have been successfully employed in the studies of the RG flow of Quantum Chromodynamics and also in the context of Condensed-matter Physics~\cite{Dupuis:2020fhh}, obtaining results in agreement with those derived from conformal-bootstrap techniques~\cite{Balog:2019rrg}. In the context of gravity, the FRG has been extensively used to assess the existence of the so-called Reuter fixed point and investigate its features in progressively larger truncations of the full theory space\footnote{Let us remark that, although additional (ghost) poles appear upon truncating a derivative expansion of the effective action, they do not necessarily spoil unitarity, since they could be truncation artifacts~\cite{Platania:2020knd}.}. Despite these studies provide strong evidence for the existence of the Reuter fixed point (see \cite{Percacci:2017fkn,Reuter:2019byg} and references therein), a complete proof is still lacking due to the practical necessity to truncate. To this end, frameworks such as tensor models~\cite{Eichhorn:2017xhy,Eichhorn:2019hsa} and dynamical triangulations~\cite{Ambjorn:2012jv,Laiho:2017htj,Loll:2019rdj} could provide alternative ways to seek asymptotic safety in quantum gravity~\cite{Ambjorn:2011cg,Laiho:2011ya,Laiho:2016nlp,Ambjorn:2016cpa,Platania:2017djo,Knorr:2018kog,Ambjorn:2018qbf}, thus complementing the FRG and circumventing its limitations.

On the other hand, string theory appears to provide a consistent UV completion, but the resulting all-order corrections are best understood insofar as supersymmetry remains unbroken. Aside from peculiar settings, in general only the computation of the first few corrections is currently feasible. In the context of (perturbative) string theory, corrections are organized in a double expansion in the string coupling $g_s$ and the ``Regge slope'' $\alpha'$, which defines the string length scale $\ell_\text{s} \equiv \sqrt{\alpha'}$. At least within the first few orders in $g_s$, string theory provides a well-defined recipe to compute curvature corrections,  weighed by $\alpha'$, in a systematic fashion, to arbitrarily high orders in $\alpha'$. In particular, to leading order in $g_s$ they can be determined from the conditions that the two-dimensional quantum field theory describing the string worldsheet embedded in a generic $D$-dimensional spacetime be free of Weyl anomalies~\cite{Fradkin:1985fq,Callan:1985ia,Callan:1986jb}. The simplest solutions to this consistency condition entail that the spacetime dimension $D$ be the critical dimension, namely $26$ for the bosonic string~\cite{Polyakov:1981rd} and $10$ for the superstring~\cite{Polyakov:1981re}, while ``non-critical'' backgrounds involve the presence of strong curvatures and/or large fluxes. However, even in the best-understood critical case the computation of all-order curvature corrections appears unfeasible at present, with the exception of particularly simple classes of backgrounds~\cite{Amati:1988ww, Amati:1988sa}.

While these peculiar classes do not include cosmological backgrounds, which are among the most relevant for phenomenological applications, in these settings curvature corrections are dramatically constrained by T-duality~\cite{Meissner:1991zj, Meissner:1996sa, Hohm:2015doa, Hohm:2019ccp, Hohm:2019jgu}, a genuinely stringy effect. The resulting mini-superspace effective actions involve a single function of the Hubble parameter, whose perturbative expansion ought to match the $\alpha'$ expansion order by order. Recent works have investigated some phenomenological aspects of effective actions of this type~\cite{Wang:2019dcj, Bernardo:2019bkz, Bernardo:2020zlc, Nunez:2020hxx}. This remarkable result provides our starting point: combining the consistency constraints of string theory with FRG methods can potentially shed light on string theory at large curvature.

In this paper we move the first step in this direction, studying $\alpha'$-corrections in cosmological backgrounds via FRG techniques. We start with a brief overview of the necessary background on string theory (in Section~\ref{sec:string_bg}) and on the FRG (in Section~\ref{sec:frg_bg}). In Section~\ref{sec:t-duality_frg} we impose invariance under T-duality and work directly with the constrained spacetime effective action within a mini-superspace ansatz~\cite{Hawking:1983hn, Hawking:1983hj, Vilenkin:1994rn, Ashtekar:2011ni, Bojowald:2015iga}, deriving flow equations and studying their solutions. In Section~\ref{sec:non-anal}, we show the existence of an exact solution to the flow equations in any spacetime dimension. Within this solution, the RG running of the Newton coupling matches {precisely} the one found in~\cite{Bonanno:2000ep} in the context of asymptotically safe gravity. Nonetheless, due to the non-minimal coupling between the dilaton and the metric degrees of freedom appearing in the Hohm-Zwiebach effective action, this solution turns out to be non-analytic and phenomenologically unviable, since it cannot reproduce the Einstein-Hilbert action at any scale. In order to investigate the existence of phenomenologically viable solutions, in Section~\ref{sec:eps_exp} we study the existence of analytic solutions to the flow equations. To this end, we first perform an $\epsilon$-expansion around two spacetime dimensions, along the lines of~\cite{1979W,Gastmans:1977ad,Christensen:1978sc,Kawai:1989yh,Kawai:1992np,Kawai:1993mb,Kawai:1995ju,Aida:1996zn}, finding a closed-form solution to leading order in $\epsilon$ without truncations, whose UV-behavior is governed by a non-Gaussian fixed point (NGFP) with a positive value $g_\ast>0$ of the (dimensionless) Newton coupling. Let us remark that the focus of our work is not the search of a UV NGFP, since string theory is expected to be UV-complete. Rather, we use the UV NGFP to determine, via the FRG, the infinitely many $\alpha'$-corrections to the effective action in terms of the finitely many relevant couplings in the UV. Although the corresponding effective action affords a closed-form expression to all orders in $\alpha'$, the resulting cosmology does not appear to deviate qualitatively from its classical counterpart. In particular, the absence of de Sitter solutions appears to support Swampland conjectures in this setting~\cite{Obied:2018sgi, Ooguri:2018wrx, Garg:2018reu} (see~\cite{Palti:2019pca} for a recent review). In Section~\ref{sec:truncations} we study the extension of this solution to higher spacetime dimensions, and in particular to $D=4$. Within polynomial truncations of the full theory space up to $\mathcal{O}(H^{10})$ in the Hubble parameter $H$, it appears that the analytic solution found in $(2+\epsilon)$ dimensions does not persists in higher dimensions: the fixed-point value of the dimensionless Newton coupling $g_\ast$ vanishes at a ``critical'' dimension $D_{crit}\simeq 2.8$ and turns negative for $D>D_{crit}$. 
If these results keep holding in higher-order truncations, our findings might indicate at least one of the following:
\begin{itemize}
\item The NGFP is highly non-perturbative in $D=4$  and cannot be detected via a truncated derivative expansion of the EAA;
\item The mini-superspace approximation is unable to capture all the relevant physics;
\item $\alpha'$-corrections are insufficient to achieve both a well-defined continuum limit and a phenomenologically viable IR limit and thus $g_s$-corrections might be crucial in this respect.
\end{itemize}
In Section~\ref{sec:phys} we discuss some potential cosmological implications of our findings, and in particular the possible existence of de Sitter solutions with large curvature. We conclude in Section~\ref{sec:conclusions} with a summary and a discussion of our results.

\section{Background on string theory}\label{sec:string_bg}    

Let us begin with a brief overview of some results of string theory that we shall make use of in the following sections. Specifically, in Section~\ref{sec:ws_bg} we outline how spacetime gravitational effective actions emerge from the worldsheet formulation of string perturbation theory, and in Section~\ref{sec:t-duality_bg} we describe how the resulting mini-superspace cosmological actions are heavily constrained by T-duality. In the following we shall use the ``mostly-plus'' signature for the spacetime metric, and we shall keep this convention in the remainder of this paper.
    
\subsection{The worldsheet formulation}\label{sec:ws_bg}

The worldsheet formulation of (perturbative) string theory rests on an action principle that governs the classical dynamics of a string probing a background spacetime according to an embedding of the type
\begin{eqaed}\label{eq:string_embedding}
	j : \left(\sigma \, , \tau \right) \; \mapsto \; X^\mu(\sigma, \tau) \, .
\end{eqaed}
where $(\tau,\sigma)$ are local coordinates on the worldsheet. $X^\mu(\tau,\sigma)$ then sweeps the string worldsheet in the background spacetime. The natural starting point is the Nambu-Goto action, which takes the simple form
\begin{eqaed}\label{eq:nambu-goto}
	S_\text{NG} = - \, \frac{1}{2\pi \alpha'} \int d\sigma \, d\tau \, \sqrt{- \det j^*G}
\end{eqaed}
in terms of the background metric $G$ and the pull-back $j^*$ of the embedding. A perturbative quantization of this theory can be carried out introducing an auxiliary dynamical metric $\gamma$ on the worldsheet, obtaining the Polyakov action~\cite{Polyakov:1981rd}
\begin{eqaed}\label{eq:polyakov}
	S_\text{P} = - \, \frac{1}{4\pi \alpha'} \int d\sigma \, d\tau \, \sqrt{- \, \det \gamma } \, G_{\mu \nu}(X) \, \gamma^{\alpha \beta} \, \partial_\alpha X^\mu \, \partial_\beta X^\nu \, .
\end{eqaed}
The action of Eq.~\eqref{eq:polyakov} takes the form of a non-linear $\sigma$-model whose target spacetime is described by the metric $G_{\mu\nu}(X)$, with the important difference that $\gamma$ is dynamical. Remarkably, the resulting spectrum in flat spacetime contains massless states associated to the (quanta of the) gravitational field, the Kalb-Ramond $2$-form $B_2$ and the dilaton $\phi$. Using the state-operator correspondence one can show that introducing the corresponding coherent states is tantamount to modifying the spacetime metric $G$. More generally, coherent states of this type are related to the couplings of the most general (renormalizable) two-dimensional $\sigma$-model, which is described by~\cite{Callan:1985ia}
\begin{eqaed}\label{eq:polyakov_full}
	S_\sigma = - \, \frac{1}{4\pi \alpha'} \int d\sigma \, d\tau \, \sqrt{- \, \det \gamma } \, & \bigg[ G_{\mu \nu}(X) \, \gamma^{\alpha \beta} \, \partial_\alpha X^\mu \, \partial_\beta X^\nu \\
	& + B_{\mu \nu}(X) \, \epsilon^{\alpha \beta} \, \partial_\alpha X^\mu \, \partial_\beta X^\nu \\
	& + \alpha' \, R^{(2)} \, \phi(X) \bigg]
\end{eqaed}
up to the inclusion of fermions, where $R^{(2)}$ is the worldsheet Ricci scalar. Moreover, the introduction of the auxiliary worldsheet metric $\gamma$ entails the presence of Weyl invariance, which acts according to
\begin{eqaed}\label{eq:weyl_action}
	\gamma_{\alpha \beta} \; \mapsto \; \Omega^2(\sigma , \tau) \, \gamma_{\alpha \beta}
\end{eqaed}
and is generally broken by the trace anomaly\footnote{Notice that Eq.~\eqref{eq:trace_anomaly} is written in Euclidean signature for later convenience.}
\begin{eqaed}\label{eq:trace_anomaly}
	\sqrt{\gamma} \, \langle {T^\alpha}_\alpha \rangle = - \, \Omega \, \frac{\delta W}{\delta \Omega}\bigg|_{\Omega = 1}
\end{eqaed}
in the quantum theory, where $W=\log Z$ is the generating functional of connected diagrams. More precisely, the inclusion of the dilaton breaks Weyl invariance at the classical level, but the corresponding anomaly is of order $\alpha'$ and thus combines with its quantum counterpart. Preserving Weyl invariance is therefore paramount to consistency and to a sensible geometric interpretation of the theory. In addition, the cancellation of the trace anomaly entails effective spacetime equations of motion for the background fields. Remarkably, these equations stem from spacetime effective actions that include gravity coupled to matter. Including worldsheet fermions~\cite{Polyakov:1981re} leads to the appearance of additional degrees of freedom in spacetime, both bosonic and fermionic, and most importantly of the gravitino. However, in this paper we shall focus on the simpler bosonic case. Furthermore, since the Polyakov action introduces worldsheet diffeomorphism invariance, one ought to take into account the corresponding Faddeev-Popov ghosts, whose contribution to the trace anomaly of Eq.~\eqref{eq:trace_anomaly} is proportional to the worldsheet Ricci scalar $R^{(2)}$. It combines with a similar contribution arising from the other worldsheet fields, leading to the background-independent term
\begin{eqaed}\label{eq:ghost_anomaly}
	\langle {T^\alpha}_\alpha \rangle_{\text{gh}} = \frac{D - D_\text{crit}}{24 \pi} \, R^{(2)} \, ,
\end{eqaed}
where the \textit{critical dimension} is $D_\text{crit} = 26$ in the bosonic case, while $D_\text{crit} = 10$ when one includes worldsheet fermions and, correspondingly, worldsheet supersymmetry. Therefore, since the additional contributions to $\langle {T^\alpha}_\alpha \rangle$ vanish in Minkowski backgrounds, the latter are only allowed for $D = D_\text{crit}$, and thus string perturbation theory is anomaly-free in this case. In any dimension, the resulting tree-level spacetime effective action contains a ``universal'' gravitational sector which takes the form~\cite{Callan:1985ia}
\begin{eqaed}\label{eq:string_eft}
	S_{\text{string}} \supset \frac{1}{16\pi G_\text{N}} \int d^D x \sqrt{-g} \, e^{-2\phi} \left( R + 4 \left(\partial\phi \right)^2 - \frac{1}{12} \, H_3^2 - \frac{2(D - D_\text{crit})}{3 \alpha'}\right)
\end{eqaed}
at leading order in $\alpha'$ in the ``string frame'', with $H_3 \equiv dB_2$. One can recast it in the more conventional Einstein frame with a Weyl rescaling. It is worth noting that the final term in Eq.~\eqref{eq:string_eft}, pertaining to \textit{non-critical} strings, would dominate the other terms in the $\alpha'$ expansion, in the absence of an additional expansion parameter. Achieving perturbative control of non-critical strings is therefore especially challenging, but for special backgrounds, such as those with a linear dilaton or those studied in~\cite{Amati:1988ww, Amati:1988sa}, one can show that the trace anomaly cancels exactly in $\alpha'$. However, in general one expects that non-critical strings require all-order $\alpha'$-corrections or the presence of additional expansion parameters\footnote{In critical string models where supersymmetry is broken at the string scale, similar contributions to the spacetime effective action arise. In these settings, large fluxes appear to provide suitable expansion parameters~\cite{Mourad:2016xbk}. In the super-critical case, $D>D_\text{crit}$, one can also consider large dimensions $D$~\cite{Maloney:2002rr}.}. Therefore, in the following we shall not consider the last term of Eq.~\eqref{eq:string_eft} when speaking of the tree-level effective action, instead including the cosmological term in the full $\alpha'$-corrected action.

Higher-derivative $\alpha'$-corrections to Eq.~\eqref{eq:string_eft} can be derived following the same procedure, at least in principle, while string-loop corrections, which we shall neglect in this paper, are subtler, and are weighed by (powers of the) local string coupling $g_s \equiv e^{\phi}$. Let us remark that, in general, $g_s$ is not a free parameter in string theory, since it is determined by the (vacuum value of the) dilaton. Consistently with more familiar notions of effective action, scattering amplitudes computed from worldsheet correlators of vertex operators~\cite{Friedan:1985ge} match the ones computed from the spacetime effective action in the low-energy limit, and ought to do so to all orders in $\alpha'$ and $g_s$.

\subsection{Elements of T-duality}\label{sec:t-duality_bg}

At the level of (perturbative) string theory, T-duality manifests itself whenever strings propagate on backgrounds with non-trivial 1-cycles around which strings can wind. The resulting massive excitations feature the usual Kaluza-Klein oscillations and novel winding modes, and the spectrum exhibits a discrete symmetry under their exchange, provided that the size $R$ of the relevant cycle can be ``inverted'' according to
\begin{eqaed}\label{eq:t-dual_inversion}
	R \; \leftrightarrow \; \frac{\alpha'}{R} \, .
\end{eqaed}
More generally, in the presence of a Killing isometry the proper inversion is described in detail by Buscher's rules~\cite{Buscher:1987sk, Buscher:1987qj}. Since T-duality invariance is an inherently stringy notion, its presence cannot be captured by standard tree-level effective actions\footnote{Going beyond local field theory in spacetime, it is possible to capture the low-energy dynamics in a T-duality-invariant fashion via double field theory~\cite{Tseytlin:1990nb, Berman:2007xn, Hull:2009mi, Hohm:2010pp, Copland:2011wx, Nibbelink:2012jb, Aldazabal:2013sca}, albeit its ability to fully encode $\alpha'$-corrections is still unsettled~\cite{Hronek:2020xxi, 2020arXiv201215677C, Hassler:2020wnp}.}, which encompass the low-energy dynamics in which winding modes decouple. However, in particularly simple backgrounds, curvature corrections can be heavily constrained on grounds of T-duality alone. Indeed, at the level of the low-energy effective action, taken in $D = d + 1$ spacetime dimensions for the sake of generality, T-duality appears as an internal $O(d,d,\mathbb{R})$ symmetry that acts on the universal low-energy field content, which comprises the dilaton $\phi$, the metric $G$ and the Kalb-Ramond two-form $B_2$. It has been shown~\cite{Veneziano:1991ek, Meissner:1991zj, Meissner:1996sa, Hohm:2015doa, Hohm:2019ccp, Hohm:2019jgu} that, for a time-dependent ansatz of the form
\begin{eqaed}\label{eq:time-dep_ansatz}
	ds^2 & = - \, n^2(t) \, dt^2 + h_{ij}(t) \, dx^i dx^j \, , \\
	B_{ij} & = b_{ij}(t) \, , \\
	\phi & = \phi(t) \, ,
\end{eqaed}
where $n(t)$ is the lapse function and $h$ is the metric on spatial slices, all curvature corrections can be captured by a single function of the $O(d,d,\mathbb{R})$-covariant matrix $\dot{\mathcal{S}}$, where
\begin{eqaed}\label{eq:s_matrix}
	\mathcal{S} \equiv \mqty(b \, h^{-1} & h - b \, h^{-1} \, b \\ h^{-1} & - \, h^{-1} \, b) \, ,
\end{eqaed}
while the dilaton appears, along with the spatial determinant of the metric, in the $O(d,d,\mathbb{R})$- invariant combination
\begin{eqaed}\label{eq:big_phi}
	e^{-\Phi} \equiv \sqrt{\det(h_{ij})} \, e^{-2\phi} \, .
\end{eqaed}
Leaving $\Phi(t)$ as an independent field, we shall focus on the more specific cosmological ansatz
\begin{eqaed}\label{eq:cosmological_ansatz}
	ds^2 & = - \, n^2(t) \, dt^2 + e^{2\sigma(t)} \, d\mathbf{x}^2 \, , \\
	B_2 & = 0 \, , \\
	\Phi & = \Phi(t) \, ,
\end{eqaed}
for which the tree-level action reduces to
\begin{eqaed}\label{eq:red_action}
	S_{\text{red}} = \frac{\text{Vol}_d}{16\pi G_{\text{N}}}\int dt \, \frac{1}{n} \, e^{-\Phi} \left( - \, \dot{\Phi}^2 + d \, \dot{\sigma}^2 \right) \, ,
\end{eqaed}
where $\text{Vol}_d$ is the (unwarped) volume of the $d = D-1$ spatial dimensions, $G_{\text{N}}$ is the $D$-dimensional Newton constant, and $H\equiv\dot{\sigma}$ is the Hubble parameter. Restricting to field configurations of this type defines the ``mini-superspace'' approximation~\cite{Hawking:1983hn, Hawking:1983hj, Vilenkin:1994rn, Ashtekar:2011ni, Bojowald:2015iga}. Within this framework, multi-trace and single-trace contributions in $\dot{\mathcal{S}}$ combine into a single function of $\dot{\sigma}$, and T-duality acts changing the sign of $\sigma$. Therefore, the effective action is to be even in $\sigma$ on consistency grounds. Including perturbative curvature corrections, the resulting action can then be written as an asymptotic series of the form
\begin{eqaed}\label{eq:alpha_corrections}
	S_{\text{HD}} \sim \frac{\text{Vol}_d}{16\pi G_{\text{N}}}\int dt \, \frac{1}{n} \, e^{-\Phi} \left( - \, \dot{\Phi}^2 + d \, n^2 \, \sum_{m=0}^\infty a_m\, {\alpha'}^{m-1} \left(\frac{\dot{\sigma}}{n} \right)^{2m} \right)  \, ,
\end{eqaed}
up to integration by parts and terms that vanish on-shell. The coefficients $a_m$ are related to the coefficients $c_m$ of~\cite{Hohm:2015doa, Hohm:2019ccp, Hohm:2019jgu} according to $a_m = 2 \, (-4)^m \, c_m$. At present, the extent to which the corrections of Eq.~\eqref{eq:alpha_corrections} can encode non-perturbative physics is unclear\footnote{See~\cite{Krishnan:2019mkv} for a discussion in the context of de Sitter solutions.}. While it is conceivable that worldsheet instantons could be included resumming the asymptotic series, \textit{e.g.} via FRG techniques, focusing on cosmological configurations could neglect contributions arising from (functional) traces over degrees of freedom ``orthogonal'' to mini-superspace. At any rate, our considerations rest on parametrically small string couplings $g_s = e^\phi \ll 1$, a condition that is to be verified \textit{a posteriori} within solutions of the effective equations of motion. Nonetheless, it is important to stress that the Hohm-Zwiebach classification of curvature corrections holds only for homogeneous and isotropic backgrounds. Were it more general, one could evaluate the effective action on more general configurations, including compactifications, and study the resulting fluctuations, including Kaluza-Klein and winding modes. As a final remark, let us emphasize that the corrections contained in Eq.~\eqref{eq:alpha_corrections} also encode the contribution of higher-spin massive string modes -- a key feature of string theory. These modes are integrated out to obtain the low-energy effective action for the massless modes.

\subsubsection{Wick rotation and gauge-fixing}\label{sec:wick_rotation}

In the tree-level spacetime effective action of Eq.~\eqref{eq:string_eft}, the sign in front of the dilaton kinetic term is positive. This is not problematic, since the corresponding Einstein-frame action reads
\begin{eqaed}\label{eq:einstein_eft}
	S_{\text{Einstein}} = \frac{1}{16\pi G_\text{N}} \int d^D x \, \sqrt{-g} \left( R - \frac{4}{D-2} \left(\partial\phi \right)^2 - \frac{1}{12} \, e^{-\frac{8}{D-2}\,\phi} \, H_3^2 \right) \, .
\end{eqaed}
From the Einstein-frame action of Eq.~\eqref{eq:einstein_eft} one can observe that the usual Wick rotation $t = -i \, t_E$ leads to a positive-definite\footnote{Let us recall that these equations are correct when written adhering to the ``mostly-plus'' convention.} kinetic term in the Euclidean action $S_E$, which is defined by $iS_{\text{Einstein}} = - \, S_E$. Performing this Wick rotation on the mini-superspace action of Eq.~\eqref{eq:red_action}, the (string-frame) Euclidean action defined by $S_E = -i \, S_{\text{red}}$ takes the same form and signs as $S_{\text{red}}$, since in the Einstein frame all kinetic terms become positive. The same result is achieved performing the Wick rotation on the full gravitational action, since the Ricci scalar does not change sign. Upon specializing the Wick-rotated covariant action to a cosmological ansatz, one obtains again Eq.~\eqref{eq:red_action}.

As a final remark, let us observe that, within the mini-superspace framework that we have described, fixing $n = 1$ in the (Euclidean) path integral amounts to computing
\begin{eqaed}\label{eq:faddeev-popov_trick}
	\frac{1}{\text{Vol}(\text{gauge})} \int \mathcal{D} n \, \mathcal{D} \sigma \, \mathcal{D} \Phi \, e^{- S_E} \, \Delta_{\text{FP}} \, \delta(n - 1) \, ,
\end{eqaed}
where the gauge-fixing $\delta$-functional yields a trivial Faddeev-Popov determinant $\Delta_{\text{FP}}$, since $n$ is a Lagrange multiplier. Hence, since the gauge transformations of the lapse do not involve the dynamical fields, the corresponding ghosts decouple, and we shall neglect them in the following.
  
\section{Background on the FRG}\label{sec:frg_bg}

In this section we provide a brief review of the main features of the FRG equations. The FRG is a mathematical tool to study the RG flow of quantum field theories and their universality properties. The basic idea is to convert the Wilsonian shell-by-shell integration of fluctuating modes, at the level of the path integral, into a functional integro-differential equation for the so-called EAA $\Gamma_k$, $k$ being a RG scale. The action functional $\Gamma_k$ is a scale-dependent effective action, resulting from the integration of quantum fluctuations with momenta $p\gtrsim k$. Accordingly, $\Gamma_k$ interpolates smoothly between the bare action of the theory (whereby $k\to\infty$) and the ordinary effective action (whereby $k\to0$). 

Once one has specified the symmetries of the theory, along with its field content, the flow of the EAA in the theory space is  determined by a flow equation. One of the most commonly used FRG equations is the Wetterich equation~\cite{Wetterich:1992yh,Morris:1993qb,Reuter:1996cp}
\begin{eqaed} \label{eq:wetterich}
k \partial_k \Gamma_k=\frac{1}{2}\,\mathrm{STr}\left\{ \left( \Gamma_k^{(2)}+\mathcal{R}_k \right)^{-1} k \partial_k \mathcal{R}_k \right\}\,.
\end{eqaed}
Here the supertrace STr entails an integral over continuous coordinates, as well as sums over any additional internal index. The EAA $\Gamma_k[\Phi]$ is a functional of all the fields in the theory, and $\Gamma_k^{(2)}$ is its second functional derivative with respect to these fields. In the case of gauge theories, a gauge-fixing and Faddeev-Popov ghosts have to be added, and these terms also contribute to the flow of $\Gamma_k$.
The function $\mathcal{R}_k$ is an infrared (IR) regulator, which depends on the ratio between the physical momentum $p$ and the RG scale $k$. It enters $\Gamma_k$ as an effective, $k$-dependent mass term such that modes with $p\lesssim k$ are suppressed, while those with momenta $p\gtrsim k$ are integrated out. Since $\mathcal{R}_k$ acts as an effective, scale-dependent mass, it also enters the modified-inverse propagator in the right-hand-side of the Wetterich equation. Its derivative $k\partial_k \mathcal{R}_k$ induces the flow of~$\Gamma_k$. The most standard choice for the $\mathcal{R}_k$-function is the Litim regulator $\mathcal{R}_k(p^2)=(k^2-p^2)\,\theta(k^2-p^2)$~\cite{Litim:2001up}, and we shall adopt this regulator throughout this work.

Since $\Gamma_k$ interpolates between the UV and the IR, the fixed points of its RG flow, where
\begin{eqaed}
k \partial_k \Gamma_k=0\,,
\end{eqaed}
provide all possible UV completions of the theory. In most cases such fixed points are saddles  of the RG flow: only a specific subset of RG trajectories -- those lying on the UV critical manifold of a fixed point -- will attain it in the UV limit. The IR and UV behavior of the RG trajectories is thus determined by the initial conditions of the flow which, in turn, are specified by observations at low energies. Accordingly, a theory is UV-complete if its RG trajectory, uniquely identified by observations in the IR, reaches a fixed point as $k\to\infty$. The theory is said to be asymptotically free if its RG trajectory ends in a Gaussian fixed point (GFP) in the UV. If instead its UV-completion is a NGFP, \textit{i.e.}, an interacting theory, the theory is said to be  asymptotically safe. A candidate UV-completion brings along key information, namely the number of IR-relevant directions associated with the corresponding fixed point, \textit{i.e.}, the co-dimension of the corresponding UV-critical manifold. Indeed, this represents the number of free parameters of the theory and thus provides a measure of its predictivity. Denoting the couplings of a theory by $g_i$ and the corresponding beta functions by $\beta_i(g_1,g_2,\dots)$, the number of relevant directions coincides with the number of positive ``critical exponents''~$\theta_i$, defined as (minus) the eigenvalues of the stability matrix, $S_{ij}\equiv\partial_{g_i}\beta_{j}$.

The Wetterich equation, Eq.~\eqref{eq:wetterich}, cannot be solved exactly at present, and one is forced to ``truncate'' the theory space specifying an ansatz for the EAA $\Gamma_k$. Specifically, $\Gamma_k$ is generally written in terms of a derivative or vertex expansion and truncated to a certain order. The flow equation is then employed to extract the beta functions for a finite number of couplings. It is worth mentioning that the derivative expansion of the effective action relies on a ``natural'' ordering of the operators, based on their canonical mass dimensions. Thus, insofar as the flow stays perturbative, \textit{i.e.} the scaling of the couplings is well-approximated by the canonical scaling, the flow of the truncated derivative expansion ought to provide a reliable approximation to the (projection of the) exact solution to~\eqref{eq:wetterich} on the corresponding theory sub-space. In the following sections we shall employ the FRG equations and the concepts reviewed above to compute cosmological~$\alpha'$-corrections in string theory. In practice, this amounts to promoting the coefficients $a_m$ in Eq.~\eqref{eq:alpha_corrections} to functions of the RG scale $k$ and determining their IR behavior in terms of the relevant parameters in the UV. To this end, it is more convenient to introduce a function $\mathcal{F}_k$ whose power series expansion encodes the coefficients $a_m$ and thus $\alpha'$-corrections to all orders.
    
\section{T-duality, functional RG flows and string cosmology}\label{sec:t-duality_frg}    

Let us now focus on the spacetime effective action of string theory in $D=(d+1)$ dimensions, with $d$ the dimension of spatial slices. As we have discussed in Section~\ref{sec:t-duality_bg}, for the time-dependent backgrounds of Eq.~\eqref{eq:time-dep_ansatz} T-duality constrains $\alpha'$-corrections to a large extent, and for cosmological backgrounds they are encoded in a single function of the Hubble parameters, at least perturbatively to all orders in $\alpha'$. Motivated by this remarkable result, in this section we apply FRG methods to the mini-superspace effective action, taking into account the constraints of T-duality. In general, one would expect that, along the flow, the EAA deviate from exact T-duality, due to the presence of the regulator in Eq.~\eqref{eq:wetterich}. However, since the regulator vanishes in the IR, T-duality ought to be recovered in this limit, and thus it appears reasonable that truncation errors be milder compared to more traditional computations\footnote{One can also construct T-duality-invariant regulators using the matrix $\mathcal{S}$ in Eq.~\eqref{eq:s_matrix}, and we intend to investigate the resulting flow in future work.}.

\subsection{Setup}\label{sec:setup}

In order to try to capture all-order $\alpha'$-corrections, while neglecting string-loop corrections, let us consider effective actions of the form of Eq.~\eqref{eq:alpha_corrections}. According to the general procedure outlined in Section~\ref{sec:frg_bg}, one ought to promote each parameter in the action to a function of the RG scale $k$. Note that the possibility of promoting the Meissner-Hohm-Zwiebach effective action to an EAA stems from the fact that its derivation holds even with $k$-dependent coefficients, provided that $g_s$ corrections can consistently be neglected\footnote{Let us stress that this can only be self-consistent if the resulting RG-improved solutions that one studies are such that $e^{\phi} \ll 1$. This ought to be done checking whether a background configuration of interest, obtained from the effective action, yields a small (exponentiated) dilaton vacuum expectation value. Since the resulting field equations depend on the (shifted) dilaton via its derivatives only, in our setting any region of spacetime where the dilaton is bounded can be made parametrically weakly coupled shifting it by an arbitrarily large negative constant.} and that the regulator preserves T-duality. To this end, we define the dimensionless running Newton constant according to $16\pi G_k \equiv g_k \, k^{1-d}$, and thus our ansatz for the EAA takes the form
\begin{eqaed}\label{eq:EAA_ansatz}
    \Gamma_k = \text{Vol}_d \int dt \, \frac{1}{n} \, e^{-\Phi} \left[ - \, k^{d-1} \, \frac{\zeta_k}{g_k}\,  \dot{\Phi}^2 + k^{d+1} \, n^2 \, \mathcal{F}_k\left(\frac{\dot{\sigma}}{n \, k}\right) \right] \, ,
\end{eqaed}
where $\mathcal{F}_k$ is an even dimensionless function of the dimensionless ratio $x={H}/k$ of the Hubble parameter $H=\dot{\sigma}$ to the RG scale $k$. We shall henceforth gauge-fix $n = 1$. Functional traces are to be defined according to the correct inner product on the space of fields, which is inherited from the kinetic terms of the classical action. Namely, letting $\chi \equiv (\sigma \, , \, \Phi)$, given an operator $\mathcal{K}$ on field space defined by a kernel $K$ according to
\begin{eqaed}\label{eq:operator_kernel}
    \left(\mathcal{K} \chi\right) (t) \equiv \int dt' \, K(t,t') \, \chi(t') \, ,
\end{eqaed}
its trace is given by
\begin{eqaed}\label{eq:functional_trace}
    \text{STr} \, \mathcal{K} = \int dt \, e^{-\Phi(t)} \, K(t,t) \, .
\end{eqaed}
The analog of a polynomial truncation in this setting would be
\begin{eqaed}\label{eq:poly_F}
    \mathcal{F}_k(x) = \frac{d}{g_k} \sum_{n=0}^N \hat{f}_n(k) \, x^{2n} \, , \qquad \hat{f}_1(k) = 1 \, ,
\end{eqaed}
and within truncations with $N \geq 2$ the coupling $\zeta_k$ is marginal. This suggests that setting $\zeta_k = 1$ in the EAA~\eqref{eq:EAA_ansatz} should be sensible: at the level of Eq.~\eqref{eq:EAA_ansatz}, the ratio $\frac{\zeta_k}{g_k}$ is a single independent coupling which can be renamed $\frac{1}{g_k}$ without loss of generality, and the physical Newton constant can still be extracted from the small-$x$ behaviour of $\mathcal{F}_k$. Moreover, setting $\zeta_k = \zeta_*$ simplifies the flow\footnote{In practice, this amounts to setting $f_1 = \frac{1}{\zeta_*}$.}. As we shall see, within the $\epsilon$-expansion, $\zeta_k$ is again marginal, and thus setting it to a constant $\zeta_*$ appears consistent. In order to focus on the flow of $\mathcal{F}_k$ and $g_k$, we shall work in backgrounds where $\dot{\Phi}$ and the Hubble parameter $\dot{\sigma} \equiv H \equiv k \, x$ are constant.

As a final remark, note that we shall interchangeably employ the notation $\mathcal{F}_k(x)$ and
	\begin{eqaed}\label{eq:cal_to_F}
		F_k(H^2) \equiv \mathcal{F}_k\left(x = \frac{H}{k}\right) \, ,
	\end{eqaed}
since the latter is more suited to derive the flow equations that we shall present shortly, as well as for the study of truncations presented in Section~\ref{sec:truncations}.

\subsection{Non-analytic exact solution in $D$ dimensions}\label{sec:non-anal}

The flow equations for the Newton coupling and the function $\mathcal{F}_{k}$ are highly non-linear, and thereby  several solutions might exist. 
In this section we derive an exact solution to the flow equations, which actually involves a non-analytic function of $H^{2}$, \textit{i.e.} of the Ricci scalar in the mini-superspace approximation, and is valid in any spacetime dimension.

\begin{figure}
\centering{}\includegraphics[scale=0.45]{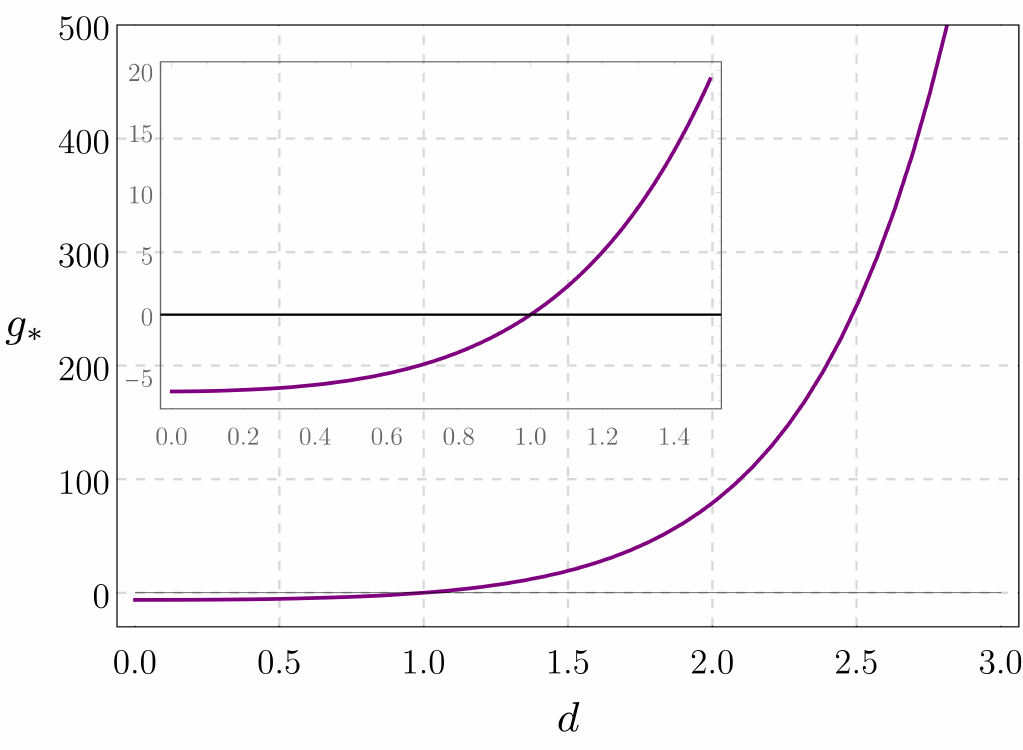}\caption{Position of the NGFP for $g_{k}$ as function of the spatial dimension
$d$.\label{Fig:exact1}}
\end{figure}

Substituting the ansatz~\eqref{eq:EAA_ansatz} for $\Gamma_k$ in the Wetterich equation~\eqref{eq:wetterich} yields the following flow equations for the (dimensionless) Newton coupling $g_k$ and the function $F_k(H^2)$
	\begin{align}
	&k\partial_{k}g_{k}=\left(d-1\right) g_k - \, \frac{2^{-d}\pi^{-1-\frac{d}{2}}}{3 \, d\,\Gamma\left(\frac{d}{2}\right)} \left(4+d+\mathcal{D}_{k}^{0}\right) g_k^2 \, , \\[0.3cm]
	& k\partial_{k}F_{k}=-\left(d+1\right)F_{k}+\frac{2^{-d}\pi^{-1-\frac{d}{2}}}{d\,\Gamma\left(\frac{d}{2}\right)\mathcal{E}_{k}}\, k\partial_{k}\mathcal{C}_{k} \nonumber \\
	& -\mathcal{C}_{k} \left[\frac{2^{2-2d}\pi^{-2-d} \left(4+d+\mathcal{D}_{k}^{0}\right)}{3 \, d^{2} \, \Gamma\left(\frac{d}{2}\right)^{2} \left(F_{k}'\right)^{2}}-\frac{2^{-d} \pi^{-1-\frac{d}{2}} (d+1)}{d \, \Gamma\left(\frac{d}{2}\right) \mathcal{E}_{k}^{2}}\right]\nonumber \\
	& -\mathcal{S}_{k}\left[ -\frac{2}{d \, \mathcal{E}_{k}} \, k\partial_{k}\mathcal{C}_{k}^{1/2} + \frac{8}{3d}\left(\frac{g_{k}^{2}F_{k}^{2}-4}{g_{k}^{2}\left(F_{k}'\right)^{4}}-\frac{2}{g_{k}\left(F_{k}'\right)^{2} \, \mathcal{E}_{k}^{2}}\right) k\partial_{k}\mathcal{C}_{k}^{3/2} \right] \label{eq:flowF} \\ 
	& -\mathcal{S}_{k} \, \mathcal{C}_{k}^{1/2} \left[ \frac{4 \, \mathcal{E}_{k}}{3 \, d^{2} \left(F_{k}'\right)^{2}} \left(\frac{6d}{g_{k}} + \frac{2^{-d} \pi^{-1-\frac{d}{2}} \left(4+d+\mathcal{D}_{k}^{0} \right)}{\Gamma\left(\frac{d}{2}\right)}\right)-\frac{d+3}{d \, \mathcal{E}_{k}}\right] \nonumber \\
	& -\mathcal{S}_{k} \, \mathcal{C}_{k}^{3/2} \left[ 
	- \, \frac{4}{3 \, d^{2} \left(F_{k}'\right)^{2} \, \mathcal{E}_{k}}\left(\frac{6d \left(d+1\right)}{g_{k}}+\frac{2^{-d}\pi^{-1-\frac{d}{2}} \left(4+d+\mathcal{D}_{k}^{0} \right)}{\Gamma\left(\frac{d}{2}\right)}\right)
	\right. \nonumber \\
	&\quad\quad\quad\quad\;\;\; \left. + \, \frac{4 \left(d+1 \right) \mathcal{E}_{k}}{d}\left(\frac{g_{k}^{2}F_{k}^{2}-4}{g_{k}^{2} \left(F_{k}' \right)^{4}}\right)
	\right] \nonumber \, ,
	\end{align}
where
\begin{align}
 & \mathcal{C}_{k}(H^{2})=\frac{F_k'(H^2)+2H^{2}F_k''(H^2)}{H^{2}}\,,\\
 & \mathcal{D}_{k}^{0}=\frac{k \partial_k F_k'}{F_k'}\bigg|_{x=0}=k\partial_k \log\left(F_k'\right)\bigg|_{x=0}\,,\\
 & \mathcal{E}_{k}(H^{2})=\sqrt{\frac{g_{k}(F_{k}')^{2}}{2g_{k}F_k(H^{2})-4}}\,,\\
 & \mathcal{S}_{k}(H^{2})=\frac{2^{-d}\pi^{-1-\frac{d}{2}} \, \text{sign}(F_{k}') \, \text{sign}(H)}{\Gamma(d/2) \, \text{sign}(\mathcal{E}_{k})}\,\text{arctanh}\left(\frac{|\mathcal{E}_{k}(H^{2})|}{\text{sign}(F_{k}') \, \text{sign}(H) \, \sqrt{\mathcal{C}_{k}(H^{2})}}\right)\,,
\end{align}
and we have dropped the dependence of various functions on $H^{2}$ for convenience. At this point, it is easy to see that
a particularly simple solution can be found imposing $\mathcal{C}_{k}=0$,
while maintaining all the other functions defined above finite. The differential
equation $\mathcal{C}_{k}(H^{2})=0$ has indeed a very simple solution, namely
\begin{equation}\label{eq:non-anal_f_sol}
F_k(H^{2})=f_{0}(k)+f_{1}(k)\sqrt{H^{2}} \, .
\end{equation}
When evaluated on this solution, all functionals defined above are
finite: one can thus safely take the limit $\mathcal{C}_{k}\to0$. In this limit the flow equations take the simple form
\begin{equation}
k\partial_{k}g_{k}=g_{k}(d-1)-\frac{2^{-d}\pi^{-1-d/2}g_{k}^{2}}{3d\,\Gamma(d/2)}\left[(4+d)+\left(\frac{k\partial_{k}f_{1}(k)}{f_{1}(k)}-1\right)\right] \, ,
\end{equation}
\begin{equation}
k\partial_{k}\left(f_{0}(k)+f_{1}(k)\sqrt{H^{2}}\right)=-(d+1)\left(f_{0}(k)+f_{1}(k)\sqrt{H^{2}}\right) \, ,
\end{equation}
and therefore one can also find an exact solution for the RG-scale dependence
of the couplings $(g_{k},f_{0}(k),f_{1}(k))$. The final solution reads
\begin{equation}
g_{k}=\frac{g_{0}g_{\ast}}{g_{\ast}k_{0}^{d-1}+g_{0}(k^{d-1}-k_{0}^{d-1})} \, k^{d-1}\label{eq:runningg} \, ,
\end{equation}
\begin{equation}\label{eq:solnaf}
F_k(H^{2})=k^{-d-1}\left(c_{0}+c_{1}\sqrt{H^{2}}\right) \, ,
\end{equation}
where $c_{0}$ and $c_{1}$ are integration constants, $g_{0}$ is
the value of the dimensionless Newton coupling at the reference scale
$k_{0}$, and the position $g_{\ast}$ of the NGFP for $g_{k}$ is given by
\begin{equation}
g_{\ast}=2^{d}d(d-1)\pi^{1+d/2}\Gamma(d/2) \, .
\end{equation}
This NGFP for $g_{k}$ yields a critical exponent $\theta_{g}=d-1$.
As expected, $g_{\ast}=0$ for $d=1$, where $\theta_{g}=0$.
For $d>1$, the NGFP emerges from the GFP and its critical exponent
$\theta_{g}$ becomes positive, indicating that for $d>1$ it is associated
with an IR-relevant direction. The variation of the position
of the NGFP is shown in Fig.~\ref{Fig:exact1}.

\begin{figure}
\includegraphics[scale=0.4]{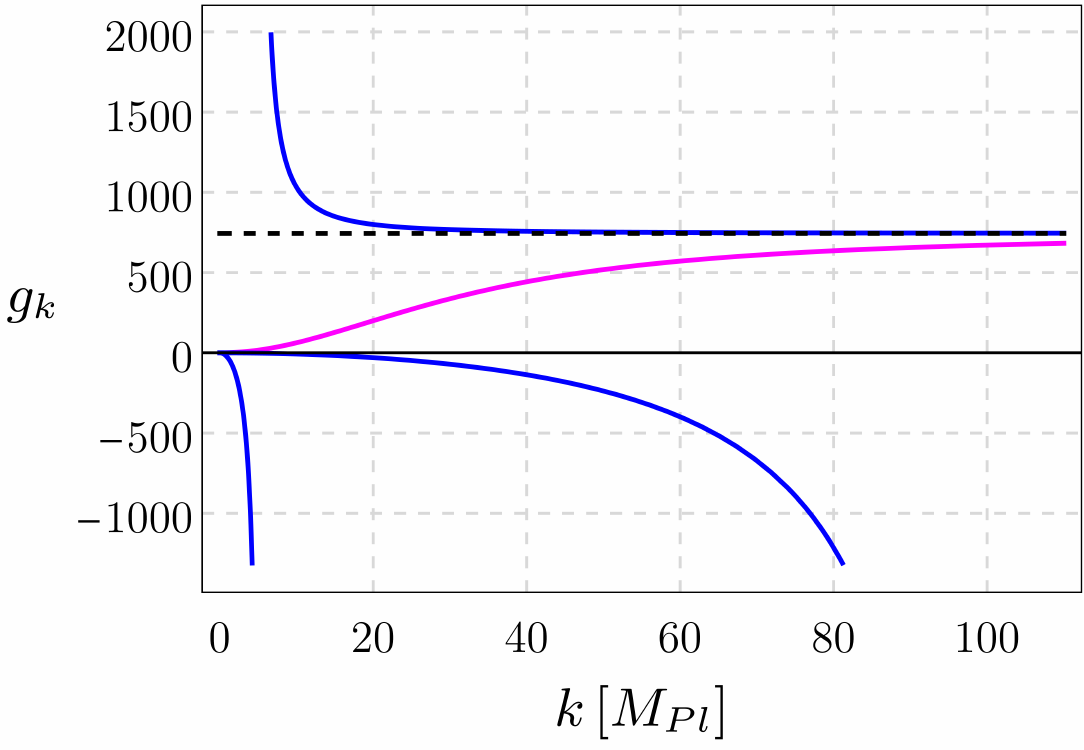}$\quad$\includegraphics[scale=0.4]{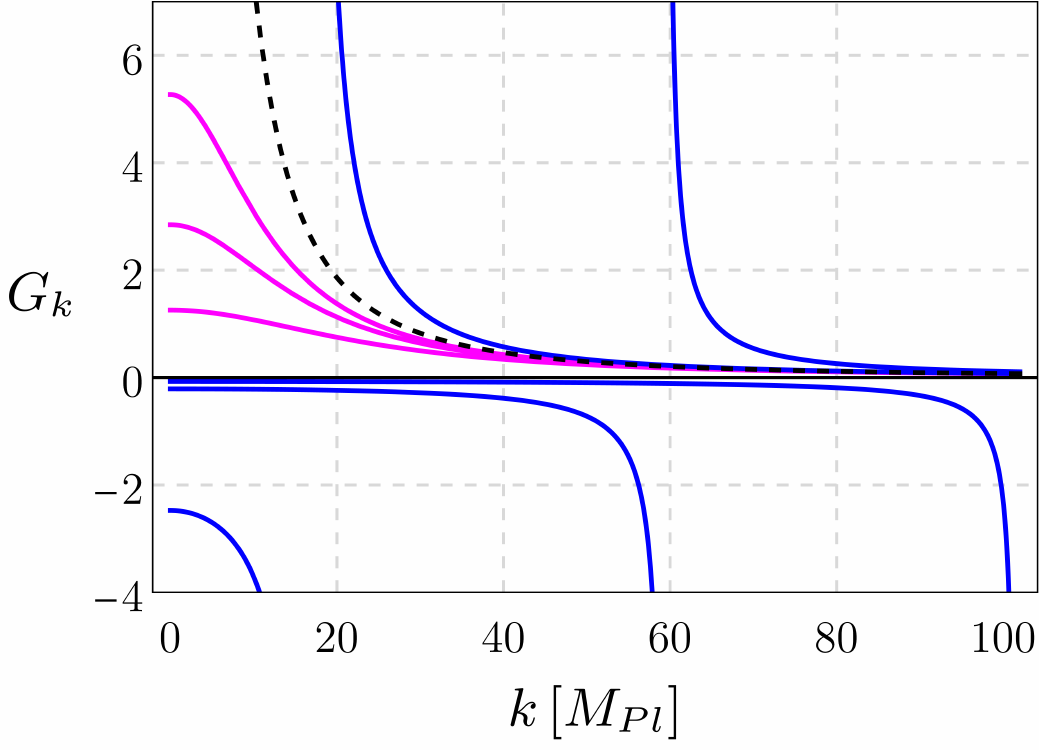}\caption{Running dimensionless (left panel) and dimensionful (right panel)
Newton coupling in $d=3$, for various initial conditions. The black-dashed
lines indicate the position of the NGFP~$g_{\ast}$. There are three
types of RG trajectories. Trajectories with $g_{k}>g_{\ast}$ are
attracted towards the NGFP at high energies but diverge in the IR.
Trajectories with $g_{k}<0$ diverge negatively at high energies and
attain the GFP in the IR. Finally, trajectories with $0<g_{k}<g_{\ast}$
interpolate between the NGFP in the UV and the GFP in the
IR. Clearly, the first two classes of trajectories (depicted
as blue lines in both plots) are unphysical, while the only phenomenologically viable trajectories are those connecting the NGFP with the GFP (magenta lines), since they allow for a finite, non-zero
dimensionful Newton coupling in the IR limit. \label{Fig:exact2}}
\end{figure}

\begin{figure}
\centering{}\includegraphics[width=0.7\textwidth]{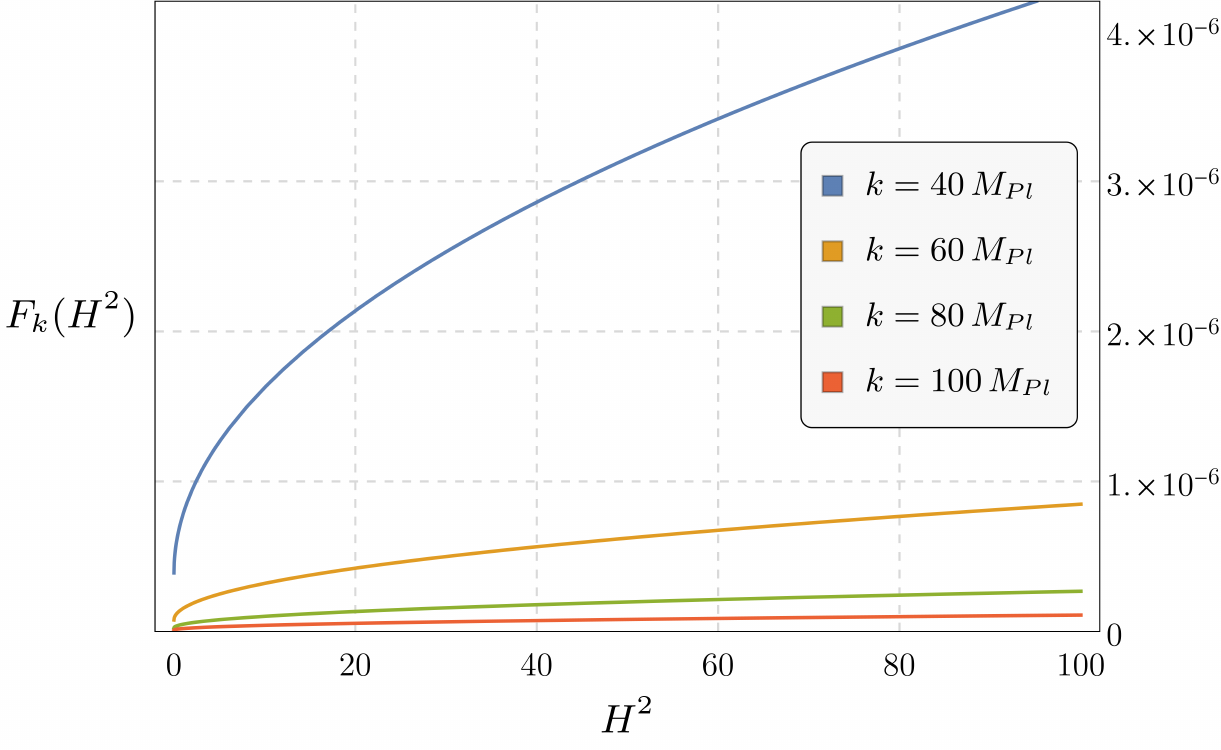}\caption{The function $F_{k}(H^{2})$ for $c_{0}=c_{1}=1$ and various values of
the RG scale $k$, in units of the Planck mass, in $d=3$ space dimensions.\label{Fig:exact3}}
\end{figure}

\begin{figure}
\includegraphics[width=0.31\textwidth]{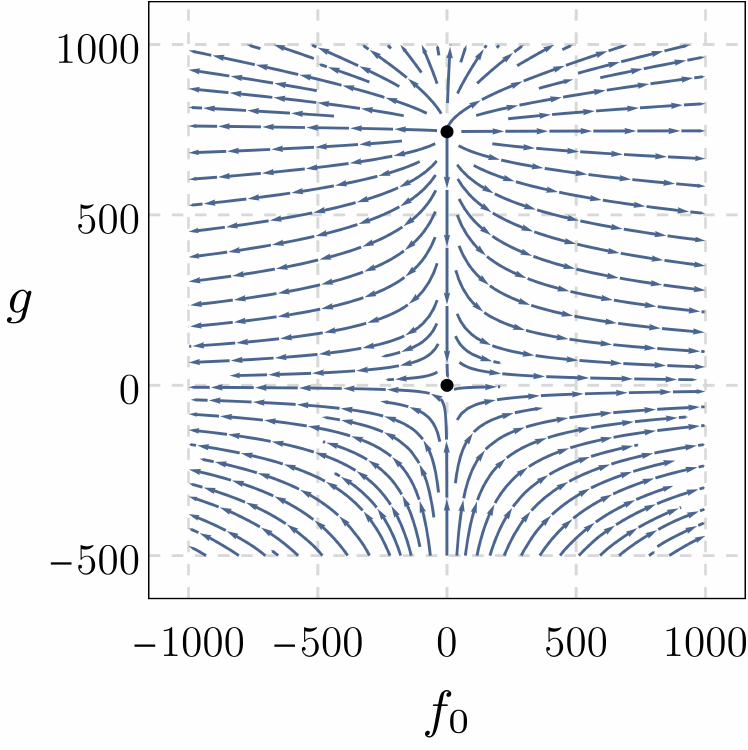}$\;\;$
\includegraphics[width=0.31\textwidth]{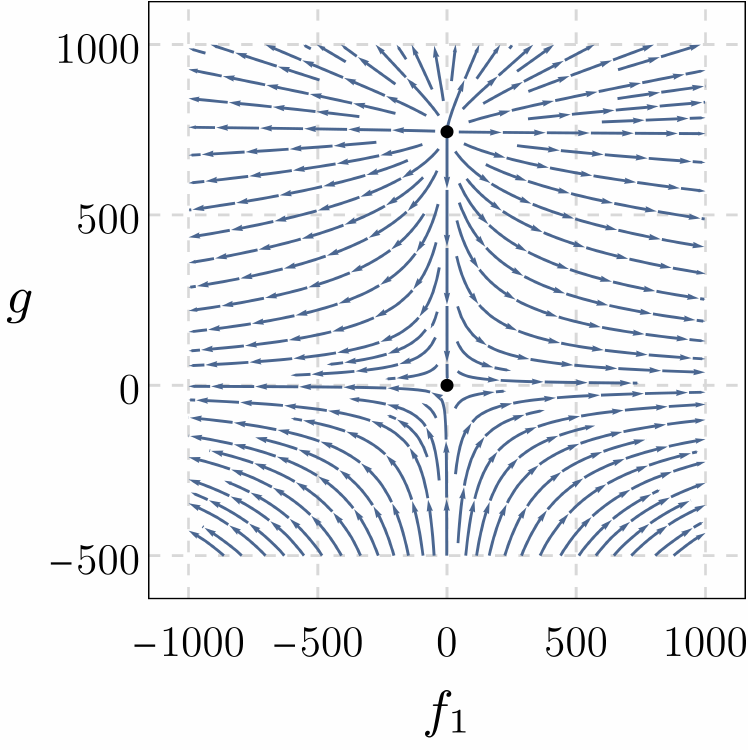}$\;\;$
\includegraphics[width=0.31\textwidth]{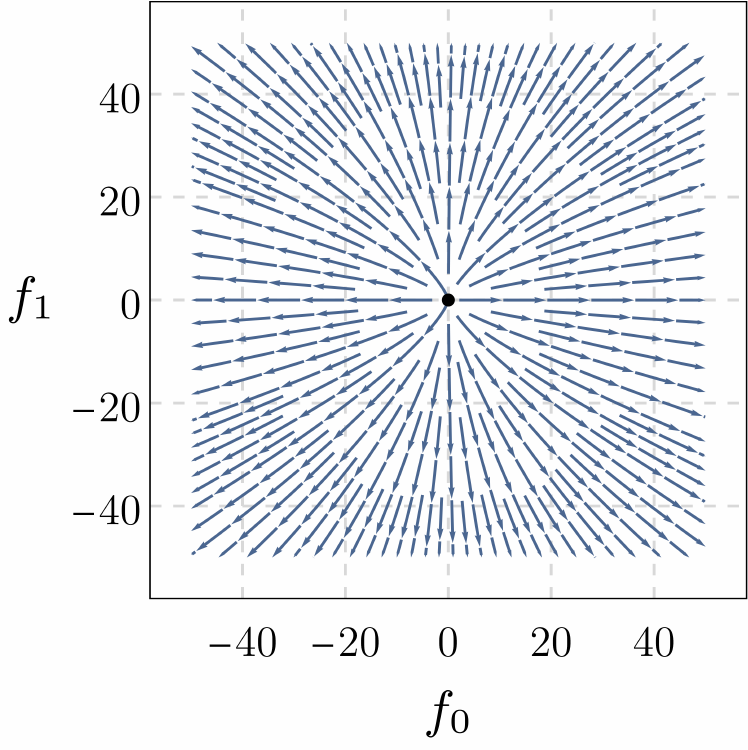}
\caption{Projection of the RG flow of $F_k$ and $g_k$ on the theory sub-spaces spanned by~$(g,f_{0})$ (left panel), $(g,f_{1})$ (central panel) and $(f_{0},f_{1})$ (right panel), for $d=3$. The Gaussian and non-Gaussian fixed points are denoted by black dots. As one can observe from the first two figures from the left, the NGFP for $g_{\ast}$ is UV-attractive and comes with three relevant directions, similarly to the Reuter fixed-point of asymptotically safe gravity. The GFP is instead a saddle point: only RG trajectories lying on the hypersurface $\{g_k=0\}$ are attracted towards the GFP in the UV. \label{Fig:exact4}}
\end{figure}

It is worth noting that the qualitative features of this NGFP for
$g_k$ match those of the Reuter fixed point encountered in the
context of asymptotically safe gravity~\cite{Bonanno:2000ep}. Moreover, at least in $D=4$, the RG-running in Eq.~\eqref{eq:runningg} matches precisely the one found in the context of asymptotically safe gravity studying the beta functions for $g_{k}$ in the absence of a cosmological constant~\cite{Bonanno:2000ep}. The corresponding dimensionful Newton coupling
\begin{equation}
G_{k}=\frac{G_{0}}{1+g_{\ast}^{-1}G_{0}(k^{d-1}-k_{0}^{d-1})} \, ,
\end{equation}
with $G_{0}=g_{0}k_{0}^{d-1}$, is indeed the same scale-dependent Newton coupling in Eq.~(2.24) of~\cite{Bonanno:2000ep}. A set of possible RG trajectories for $g_{k}$ and $G_{k}=g_{k}k^{1-d}$ is shown in Fig.~\ref{Fig:exact2}
for $d=3$.
As one can observe from the figure, in $d=3$ there exist trajectories which depart from the NGFP in the UV and reach the GFP in the IR (magenta lines). These trajectories are separatrix lines which connect the
two fixed points. In the present context only this type of trajectories
is physical, since they give a positive and finite Newton coupling in
the IR. All other trajectories (blue lines) correspond to Newton
couplings which are either negative or divergent in the IR. The flow of $F_{k}(H^2)$ of Eq.~\eqref{eq:non-anal_f_sol} takes a simple form. This function is non-analytic in~$H^{2}$, and the couplings $f_{i}(k)$ exhibit canonical scaling for all values of the RG scale $k$. The evolution of $F_{k}$ is shown in Fig.~\ref{Fig:exact3} for $c_{0}=c_{1}=1$ and $d=3$, while the projection of the full RG flow of $F_k$ and $g_k$ on various two-dimensional theory sub-spaces is depicted in Fig.~\ref{Fig:exact4}.

Let us finally remark that the solution in Eqs.~\eqref{eq:runningg} and~\eqref{eq:solnaf} is unphysical, since its IR behavior is not analytic in the curvature invariants. However, due to the non-linearity of the flow equations for $g_k$ and $F_k$, the solution we found in this section might not be unique. Indeed, in the following we shall seek (analytic) solutions, first via an $\epsilon$-expansion about $d=1$ and then via polynomial truncations. Let us note that computations based on a truncated derivative expansion of $F_k$, \textit{i.e.} on a Taylor expansion of the function $F_{k}$ about $H=0$, can only provide approximations to scaling solutions for $F_k$ which are both even and analytic in $H=0$. In other words, since the solution found in this section is non-analytic in $H^{2}$, it is not accessible by polynomial truncations.

\subsection{Analytic solution within an \texorpdfstring{$\epsilon$}{ε}-expansion}\label{sec:eps_exp}


In this section we investigate the flow equations via an $\epsilon$-expansion. To this end, for convenience we shall write all functions in terms of $x = H/k$. As we have discussed in the preceding section, the flow equation for $\mathcal{F}_k$ is considerably complicated. However, the flow equation for $g_k$ is simpler, and expressing the functional dependence on $H^2$ in terms of $x$ it reads
\begin{eqaed}\label{eq:g_flow}
	k\partial_k{g}_k = (d-1) \, g_k - \frac{d+3 + \mathcal{D}_k[\mathcal{F}_k]}{3 \cdot 2^{d+1} \, \pi^{\frac{d}{2}+1} \, \Gamma(\frac{d}{2}+1)} \, g^2_k \, ,
\end{eqaed}
where we have defined the functional
\begin{eqaed}\label{eq:ratio_functional}
	\mathcal{D}_k[\mathcal{F}_k] \equiv \frac{k \partial_k \mathcal{F}_k'}{\mathcal{F}_k'}\bigg|_{x=0}
\end{eqaed}
for later convenience. Notice that here, and in the ensuing discussion, primes denote derivatives with respect to $x$.

Assuming that at a fixed point $\mathcal{D}_k[\mathcal{F}_k]$ is well-defined\footnote{This assumption can be self-consistently checked, at least within an $\epsilon$-expansion or truncations.}, in general one finds two fixed points for $g_k$. Since for $\epsilon \equiv d-1 \ll 1$ this fixed point is $\mathcal{O}(\epsilon)$, this motivates studying the RG flow within an $\epsilon$-expansion, where the flow equation for $\mathcal{F}_k$ dramatically simplifies.

While the solution that we have described in the preceding section is exact, its low-energy behavior is phenomenologically not viable. Hence, motivated by the fact that for~$d=1$ the Newton coupling is classically marginal, in this section we investigate the RG flow of $\mathcal{F}_k$ and $g_k$ at leading order in an $\epsilon$-expansion, setting $d = 1 + \epsilon$. 

In order to properly account for the scaling of each term in the flow equations as~$\epsilon \to 0^+$, one must specify an ansatz for $g_k$ and $\mathcal{F}_k$. A consistent ansatz for $g_k$ is simply a function $g_k = \epsilon \, \gamma_k$, with $\gamma_k = \mathcal{O}(1)$, which flows from a positive fixed point~$\gamma_*$ in the UV (if $\mathcal{D}_k$ is regular in the UV) to zero in the IR. A consistent ansatz for $\mathcal{F}_k$ is then
\begin{eqaed}
\mathcal{F}_k(x) = \frac{v_k(x)}{\epsilon\, \gamma_k} + w_k(x)\,\,,
\end{eqaed}
with $v_k(x) \, , \, w_k(x) = \mathcal{O}(1)$. This ansatz is motivated by the fact that in the IR regime $\mathcal{F}_k$ is expected to adhere to the perturbative result, while the corrections are expected to be sub-leading in $\epsilon$.

Using the aforementioned ansatz for $g_k$ and $\mathcal{F}_k$, the flow equations simplify considerably and, assuming that $\mathcal{D}_k[\mathcal{F}_k]$ is non-singular\footnote{We shall verify this assumption \textit{a posteriori}.}, at leading order they take the form
\begin{eqaed}\label{eq:leading_eqs}
    & k \partial_k {\gamma}_k = \epsilon \, \frac{\gamma_k}{\gamma_*} \left( \gamma_* - \gamma_k - \frac{1}{4} \, \gamma_k \, \mathcal{D}_k[v_k] \right) \, , \\
    & \frac{k \partial_k {v}_k - x \, v_k' + 2 v_k}{\epsilon \, \gamma_k} = \mathcal{O}(1) \, ,
\end{eqaed}
where $\gamma_* = \frac{3}{2} \, \pi^2$ is the fixed-point value of $\gamma_k$. Correspondingly, the (dimensionful) Newton coupling scales as
\begin{eqaed}\label{eq:newton_coupling_UV}
  G_k \sim \frac{3\pi}{32} \, \epsilon \, k^{-\epsilon}
\end{eqaed}
in the UV. The second line of Eq.~\eqref{eq:leading_eqs} then implies that $v_k(x) = \frac{V(k \, x)}{k^2}$, where we require that the arbitrary function $V$ be analytic at the origin in order to be consistent with perturbative $\alpha'$-corrections. Furthermore, in order to allow for a UV fixed point, the function $v_k(x)$ must be at most quadratic,
\begin{eqaed}\label{eq:V_quadratic}
    v_k(x) = \frac{\Lambda}{k^2} + \frac{x^2}{\zeta_*} \,.
\end{eqaed}
Here the coefficient of $x^2$ is the marginal deformation corresponding to the wave-function renormalization of $\Phi$, as discussed in the preceding section. In addition, this solution has $\mathcal{D}_k[v_k] = 0$, so that $\mathcal{D}_k[\mathcal{F}_k] = \mathcal{O}(\epsilon)$ upon including the sub-leading term $w_k(x)$. The flow equation for $\gamma_k$ can then be solved exactly. Its solution reads
\begin{eqaed}\label{eq:gamma_flow}
    \gamma_k = \frac{\gamma_*}{1 + c \, k^{-\epsilon}}\,,
\end{eqaed}
where $c$ is an integration constant. Interestingly, this solution matches the result in Eq.~\eqref{eq:runningg} that we have found in Section~\ref{sec:non-anal} for $d=1+\epsilon$. In particular, for $c > 0$ the flow interpolates smoothly between $\gamma_* = \frac{3}{2} \, \pi^2$ in the UV and zero in the IR, where $\gamma_k$ scales as $\gamma_k \sim \frac{\gamma_*}{c} \, k^{\epsilon}$ and, correspondingly,
\begin{eqaed}\label{eq:newton_constant_IR}
    G_k \sim \frac{3\pi}{32} \, \frac{\epsilon}{c}\,.
\end{eqaed}
Since $\gamma_k$ flows very slowly, with $\dot{\gamma_k} \sim \epsilon \, \gamma_k$, one can neglect contributions involving $\dot{\gamma_k}$ in the sub-leading flow equation for $w_k$, since the latter involves also $\mathcal{O}(1)$ terms which dominate over those proportional to $k\partial_k\gamma_k$. 
Using these results and observations, the resulting flow equation for $w_k$ can be solved analytically, yielding
\begin{eqaed}\label{eq:w_flow}
    \gamma_* \, w_k(x) & = \frac{1}{2} + \left( \frac{3}{4} - \frac{\gamma_*}{\gamma_k} \right) \frac{\Lambda}{k^2} \, \log\left(\frac{k}{k_0} \right) + \frac{3\Lambda}{8k^2} \, \log\left( 4 - \frac{2\Lambda}{k^2} - \frac{x^2}{\zeta_*} \right) \\
    & + \left( \left(\frac{\gamma_*}{\gamma} - \frac{1}{2}\right) \log\left(x\right) + \frac{1}{4} \, \log\left( 4 - \frac{2\Lambda}{k^2} - \frac{x^2}{\zeta_*} \right) \right) \frac{x^2}{\zeta_*} \\
    & + \left(1 + \frac{\Lambda}{4k^2} + \frac{x^2}{4\zeta_*} \right)
    \frac{\text{arctanh}\left( B_k(x) \right)}{B_k(x)} \\
    & + \gamma_* \, \frac{W(k \, x)}{k^2} \, ,
\end{eqaed}
where we have defined the combination
\begin{eqaed}\label{eq:arc_arg}
    B_k(x) \equiv \frac{x}{\sqrt{\zeta_*} \sqrt{\frac{2\Lambda}{k^2} - 4 + \frac{2x^2}{\zeta_*}}}
\end{eqaed}
and, once more, $W$ is an arbitrary function which we choose according to
\begin{eqaed}\label{eq:W_quadratic}
    \frac{W(k \, x)}{k^2} = \frac{w_0}{k^2} + w_2 \, x^2 \, ,
\end{eqaed}
with $w_0 \, , w_2$ constants. The arbitrary scale $k_0$ can thus be shifted modifying $w_0$, while $w_2$ can be fixed by requiring that the quadratic terms reproduce the classical contribution $\frac{d \, x^2}{\zeta_* \, g_k}$ in the IR. 

The large-$x$ behavior of the resulting solution $\mathcal{F}_k = \frac{v_k}{\epsilon \, \gamma_k} + w_k$ is consistent on dimensional grounds: it corresponds, at fixed $\dot{\sigma} = k \, x$, to the small-$k$ regime, which is perturbative and ought to reconstruct the classically marginal operator $\propto x^{2+\epsilon} \propto R^{\frac{D}{2}}$. Indeed,
\begin{eqaed}\label{eq:large-x_asymp}
    \mathcal{F}_k(x) \sim \frac{x^2}{\zeta_* \, g_k} \left(1 + \epsilon \, \log\left(x\right) \right) \sim \frac{x^{2+\epsilon}}{\zeta_* \, g_k} \, .
\end{eqaed}
Using these results, one can derive the leading-order IR coefficients for higher-derivative corrections to the classical Lagrangian, which have the correct classical scaling. As a result, the dependence on $k$ disappears once the dimensionless variable $x$ is replaced by $x = \frac{H}{k}$. In detail, expanding $\mathcal{F}_k(x)$ in powers of $\frac{x}{\sqrt{\zeta_*}}$ and setting $\zeta_* = 1$, the first few corrections to the classical Lagrangian read
\begin{eqaed}\label{eq:first_corrections}
    e^{\Phi} \, \mathcal{L}_{\text{HD}} \sim \frac{29}{480\pi^2 \Lambda} \, H^4 - \frac{53}{3360\pi^2 \Lambda^2} \, H^6 + \frac{1591}{241920\pi^2 \Lambda^3} \, H^8 \, .
\end{eqaed}
These corrections are $\mathcal{O}(\epsilon)$ with respect to the ``classical'' terms, whose leading-order IR contribution to the Lagrangian takes the form
\begin{eqaed}\label{eq:classical_IR_terms}
    e^{\Phi} \, \mathcal{L}_{\text{tree}} \sim \frac{1}{16\pi G_\text{N}} \left( - \dot{\Phi}^2 + \Lambda + H^2 \right)
\end{eqaed}
at a suitable IR scale $k = k_{\text{IR}}$ such that $g_{k_{\text{IR}}} = 16\pi G_\text{N} + \mathcal{O}(\epsilon)$. This identifies the relevant deformation $\Lambda$ as the leading contribution to the low-energy cosmological constant in the \textit{string-frame}\footnote{In the Einstein frame, it takes the form of an exponential potential for the dilaton $V(\phi) = \Lambda \, e^{\frac{4}{D-2} \, \phi}$.}. The size of the quartic correction approaches that of the classical curvature term when
\begin{eqaed}\label{eq:relative_correction}
   H^2 \approx \Lambda \, ,
\end{eqaed}
from which one could be tempted to identify $\Lambda \approx \frac{1}{\alpha'}$ as an UV cutoff scale. Indeed, the precise value of $\Lambda$ is proportional to $\frac{D_\text{crit} - D}{\alpha'}$, and encodes the deviation from the critical dimension~\cite{Veneziano:1991ek, Fradkin:1984pq, Callan:1985ia}. The coefficients in Eq.~\eqref{eq:first_corrections} can also be obtained by expanding the leading-order (in $\epsilon$) IR result
\begin{eqaed}\label{eq:hd_ir_lagrangian}
    e^{\Phi} \, \mathcal{L}_{\text{HD}} \sim \frac{\Lambda}{6\pi^2} \, L\left( \frac{H^2}{\Lambda} \right)
\end{eqaed}
in powers of $H$, where
\begin{eqaed}\label{eq:L_func}
    L(s) & \equiv - 1 - \frac{23}{12} \, s + \left(\frac{3}{2} + s \right) \log\left(1 + \frac{s}{2} \right) \\
    & + \left(1+s\right)^{\frac{3}{2}} \, \sqrt{\frac{2}{s}} \, \text{arctanh}\left( \sqrt{\frac{s}{2\left(1+s\right)}} \right) \, .
\end{eqaed}
All in all, reinstating the lapse $n$, the leading-order (in $\epsilon$) IR string-frame effective action for cosmological ansatze reads
\begin{eqaed}\label{eq:string-frame_IR_EA}
    \Gamma_{\text{string}} = \frac{\text{Vol}_{1+\epsilon}}{16\pi G_{\text{N}}} \int dt \, n \, e^{-\Phi} \left[\Lambda - \frac{\dot{\Phi}^2}{n^2} + \frac{\dot{\sigma}^2}{n^2} + \frac{8}{3\pi} \, G_{\text{N}} \, \Lambda \, L\left(\frac{\dot{\sigma}^2}{n^2\,\Lambda} \right) \right] \, .
\end{eqaed}
In Section~\ref{sec:phys} we shall comment on some potential cosmological consequences of these results, assuming that similar solutions and effective actions extend to the phenomenologically relevant case of $D=4$ spacetime dimensions.

\subsection{Analytic fixed-point solutions within truncations}\label{sec:truncations}

The flow equations for $g_{k}$ and $F_{k}$
admit an exact non-analytic solution valid for any dimension $D$ (cf. Section~\ref{sec:non-anal}) and an analytic solution in $D=(2+\epsilon)$ dimensions, characterized by an UV fixed point with $g_{\ast}>0$ and a phenomenologically viable IR limit (cf. Section~\ref{sec:eps_exp}). If the latter solution extended to $D=4$, the theory would be characterized by a UV fixed point with $g_\ast>0$ and a positive cosmological constant in the IR, thus realizing a scenario similar to the one proposed in~\cite{deAlwis:2019aud}. In this section we investigate whether the analytic solution found in $D=(2+\epsilon)$ dimensions can indeed extend to $D=4$. 

Since an analytic exact solution in arbitrary dimensions $d$ cannot be found in closed form, in order to investigate the extension of the fixed point found in the previous section to $d=3$ we shall truncate the theory space choosing an ansatz for $F_k$ of the form\footnote{The condition that $F_k$ be even in $H$ simplifies the computation of beta functions, and is required by T-duality.}
\begin{equation}
F_{k}(H^{2})=\sum_{n=1}^{N}f_{2n-2}(k)\left(k^{-2}H\right)^{2n-2},
\end{equation}
with $N$ finite. For a given truncation order $N$, one can determine the beta functions for all couplings associated with (even) powers of $H$ up to the order $H^{2(N-1)}$. Specifically, once $N$ is chosen, the ansatz for $F_{k}$ is to be replaced into  Eq.~\eqref{eq:flowF}. The beta functions $\beta_{n}\equiv k\partial_k f_n(k)$ for the couplings $\{f_{0}(k),\dots,f_{2N-2}(k)\}$ can then be computed expanding the flow equation for $F_k$ about $H=0$ up to order $N$ and requiring that the coefficients of the expansion vanish. For a given $N$, one can thus investigate the fixed-point structure of the flow equations as a function of $d$. One can then investigate the robustness of the results progressively enlarging the truncation order $N$. In fact, if a fixed-point functional exists in the full theory space and is analytic, it should be possible to detect it within truncations, and its critical exponents should converge to stable values upon increasing the truncation order. Vice versa, if a fixed point disappears increasing the truncation order, or if its critical exponents do not converge, it is likely that the fixed point is spurious, \textit{i.e.}, that it is a truncation artifact rather than a genuine feature of the theory. In the following we shall apply these techniques, investigating the fixed-point structure of the truncated flow of~$F_k$ up to $N=5$, namely up to $\mathcal{O}(H^{10})$.

\begin{figure}
\centering{}\includegraphics[width=0.65\textwidth]{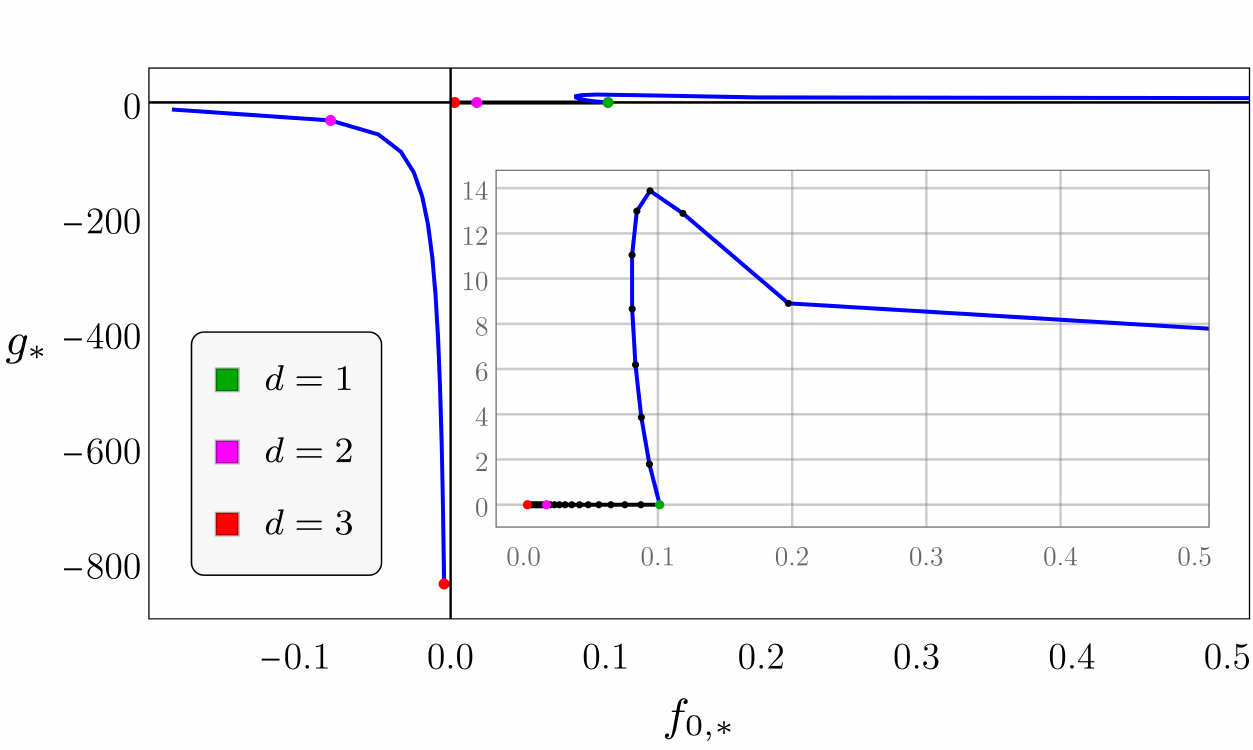}\caption{Position of the GFP and NGFP for increasing values of the spatial
dimension $d$ and $N=2$. For $d=1$ the two fixed points coincide
(green dot). Increasing $d$, the GFP and NGFP split and move apart
from each other. In $d=2$ the two fixed points are depicted by magenta
dots, while in $d=3$ they are depicted by red dots. The smaller black dots in between are the two fixed points for non-integer dimensions. The trajectories drawn by the Gaussian and non-Gaussian fixed points for increasing values of $d$ are shown as black and blue lines respectively. As can be seen from this figure, for $N=2$ the NGFP in $d=1+\epsilon$ and $d=3$ dimensions are not continuously connected: there exists a critical dimension $d_{crit}<2$ beyond which $g_{\ast}$ vanishes and becomes negative, while $f_{0,*}$ diverges at $d_{crit}$ and re-emerges at $d>d_{crit}$ with arbitrarily large
and negative values. \label{Fig:truncation1}}
\end{figure}

\begin{figure}
$\;\;$\includegraphics[scale=0.37]{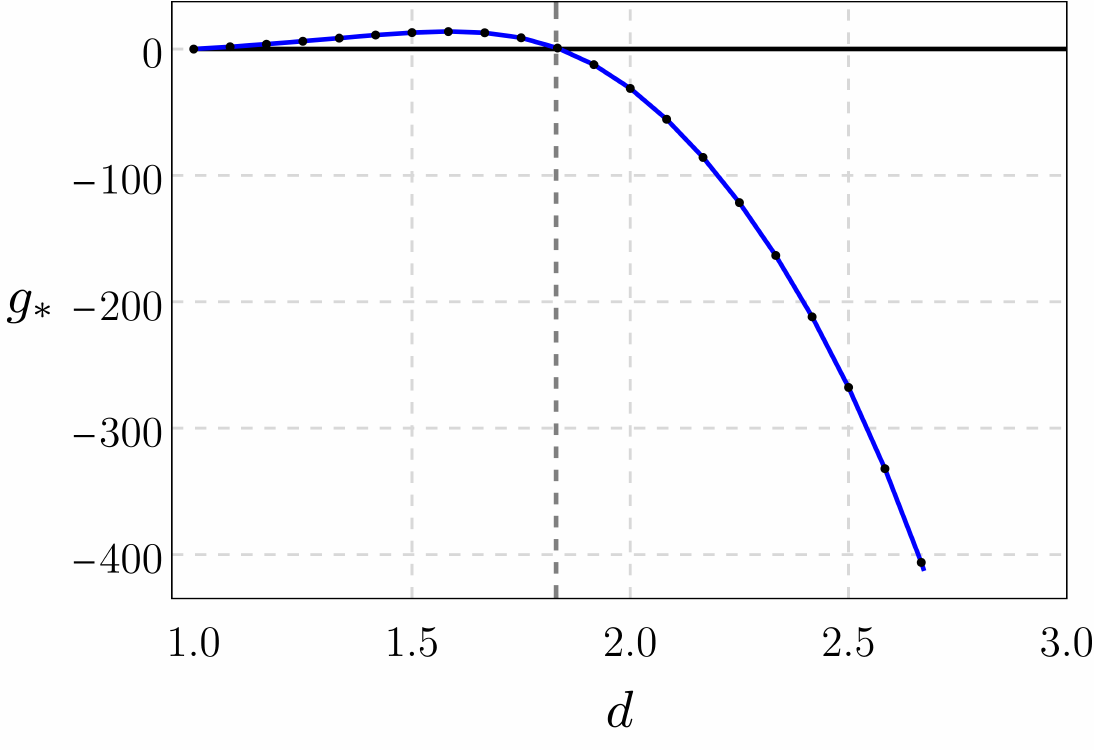}$\;\;$
\includegraphics[scale=0.37]{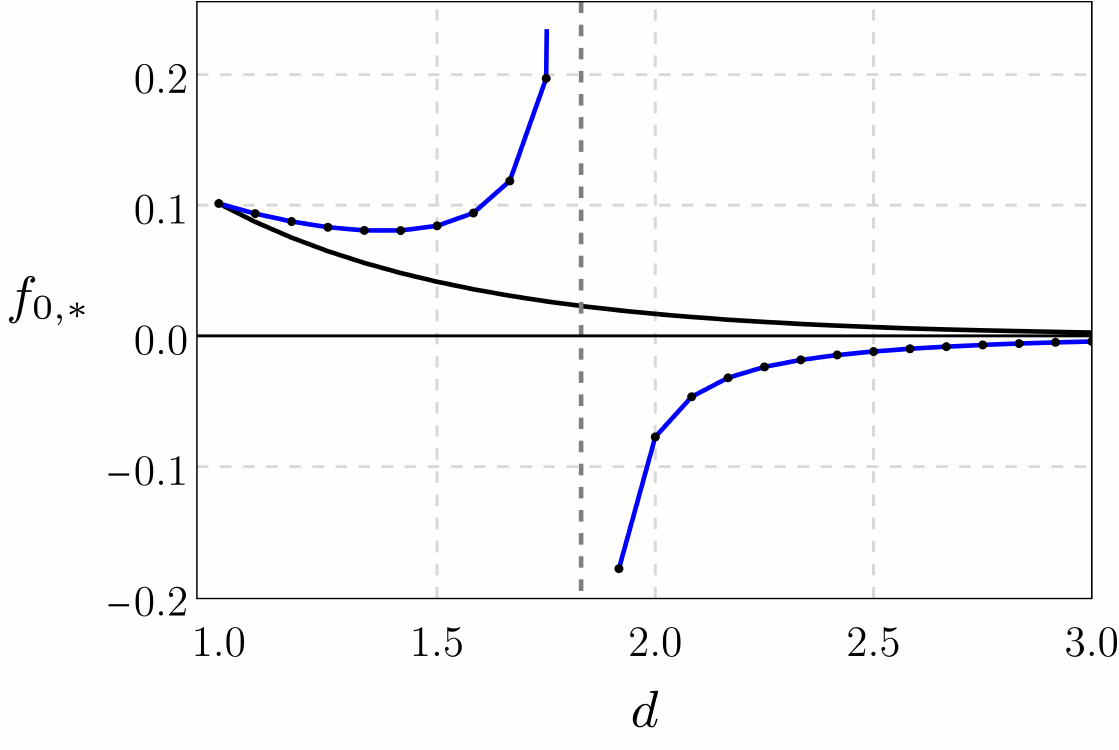}
\caption{Variation of $g_{\ast}$ and $f_{0,*}$ with the spatial dimension
$d$ (blue lines), for $N=2$. The NGFP emerging from the GFP in $d=1+\epsilon$
dimensions disappears at a critical dimension $d_{crit}<2$ (gray
dashed line) and is replaced by another fixed point at $g_{\ast}<0$
which seems to persist up to $d=3$. Both fixed points are characterized by $f_{2,*}=0$ for any $d$. The variation of the position of the GFP (black line) is also shown for comparison. \label{Fig:truncation2}}
\end{figure}

For any dimension $D$ there is a line of GFPs. The universality properties of this free fixed point are summarized in Tab.~\ref{tab:gfpD2} and Tab.~\ref{tab:gfpD4} for $D=2$ and $D=4$ respectively.\\
\begin{table}
\begin{centering}
\begin{tabular}{|c||c|c|c|c|c|c||c|c|c|c|c|c|}
\hline 
$N$ & $g_{\ast}$ & $f_{0}^{\ast}$ & $f_{2}^{\ast}$ & $f_{4}^{\ast}$ & $f_{6}^{\ast}$ & $f_{8}^{\ast}$ & \multicolumn{6}{c|}{Critical Exponents}\tabularnewline
\hline 
\hline 
\multirow{1}{*}{2} & $0$ & $\pi^{-2}$ & $\forall f_{2}$ &  &  &  & $2$ & $0$ & $0$ &  &  & \tabularnewline
\hline 
\hline 
\multirow{1}{*}{3} & $0$ & $\pi^{-2}$ & $\forall f_{2}$ & $0$ &  &  & $2$ & $0$ & $0$ & $-2$ &  & \tabularnewline
\hline 
\hline 
\multirow{1}{*}{4} & $0$ & $\pi^{-2}$ & $\forall f_{2}$ & $0$ & $0$ &  & $2$ & $0$ & $0$ & $-2$ & $-4$ & \tabularnewline
\hline 
\hline 
\multirow{1}{*}{5} & $0$ & $\pi^{-2}$ & $\forall f_{2}$ & $0$ & $0$ & $0$ & $2$ & $0$ & $0$ & $-2$ & $-4$ & $-6$\tabularnewline
\hline 
\end{tabular}
\par\end{centering}
\caption{GFP line in $D=2$. \label{tab:gfpD2}}
\end{table}
\begin{table}
\begin{centering}
\begin{tabular}{|c||c|c|c|c|c|c||c|c|c|c|c|c|}
\hline 
$N$ & $g_{\ast}$ & $f_{0}^{\ast}$ & $f_{2}^{\ast}$ & $f_{4}^{\ast}$ & $f_{6}^{\ast}$ & $f_{8}^{\ast}$ & \multicolumn{6}{c|}{Critical Exponents}\tabularnewline
\hline 
\hline 
\multirow{1}{*}{2} & $0$ & $(12\pi^{3})^{-1}$ & 0 &  &  &  & $-2$ & $4$ & $2$ &  &  & \tabularnewline
\hline 
\hline 
\multirow{1}{*}{3} & $0$ & $(12\pi^{3})^{-1}$ & 0 & $\forall f_{4}$ &  &  & $-2$ & $4$ & $2$ & $0$ &  & \tabularnewline
\hline 
\hline 
\multirow{1}{*}{4} & $0$ & $(12\pi^{3})^{-1}$ & 0 & $\forall f_{4}$ & $0$ &  & $-2$ & $4$ & $2$ & $0$ & $-2$ & \tabularnewline
\hline 
\hline 
\multirow{1}{*}{5} & $0$ & $(12\pi^{3})^{-1}$ & 0 & $\forall f_{4}$ & $0$ & $0$ & $-2$ & $4$ & $2$ & $0$ & $-2$ & $-4$\tabularnewline
\hline 
\end{tabular}
\par\end{centering}
\caption{GFP line in $D=4$, for various values of $N$. \label{tab:gfpD4}}
\end{table}
Studying the fixed-point structure of the beta functions for increasing
values of the truncation order $N$ and spatial dimension $d$, it
is in principle possible to understand whether the fixed point found analytically in $d=(1+\epsilon)$ is also present in $d=3$.
For $N=2$, the only couplings involved are $g_{k}$, $f_{0}(k)$ and $f_{2}(k)$ and all Gaussian and non-Gaussian fixed points are characterized by $f_{2,*}=0$ for all $d$. One can thus set $f_{2}(k)=0$ and visualize how $f_{0,*}$ and~$g_{*}$ for the NGFP vary with the spatial dimension $d$. The result is depicted in Fig.~\ref{Fig:truncation1}. In $d=1$ there is only the GFP. Increasing the value of $d$, a NGFP emerges from the GFP. For $d\gtrsim1$,
the NGFP is located at positive values of~$f_{0,*}$ and~$g_{*}$. Increasing $d$, this value of $g_{\ast}$ rapidly decreases, vanishes, and then
becomes negative. At the same time,~$f_{0,*}$ moves towards arbitrary large and positive values. This is shown in Fig.~\ref{Fig:truncation2}, depicting the variation of $g_{\ast}$ and $f_{0,*}$ with $d$. As is apparent from the figure, there exists a critical dimension $d_{crit}\simeq1.8$ where $g_{\ast}$ vanishes, while $f_{0,*}$ diverges. For $d>d_{crit}$, there is only a NGFP located in the unphysical part of the theory space,  $g(k)<0$. Thus, for $N=2$, the NGFP found in $d=1+\epsilon$ seems not to survive in higher dimensions: it disappears at $d=d_{crit}$
and it is replaced by another UV NGFP with $g_{*}<0$. 

\begin{figure}
$\;\;$\includegraphics[scale=0.42]{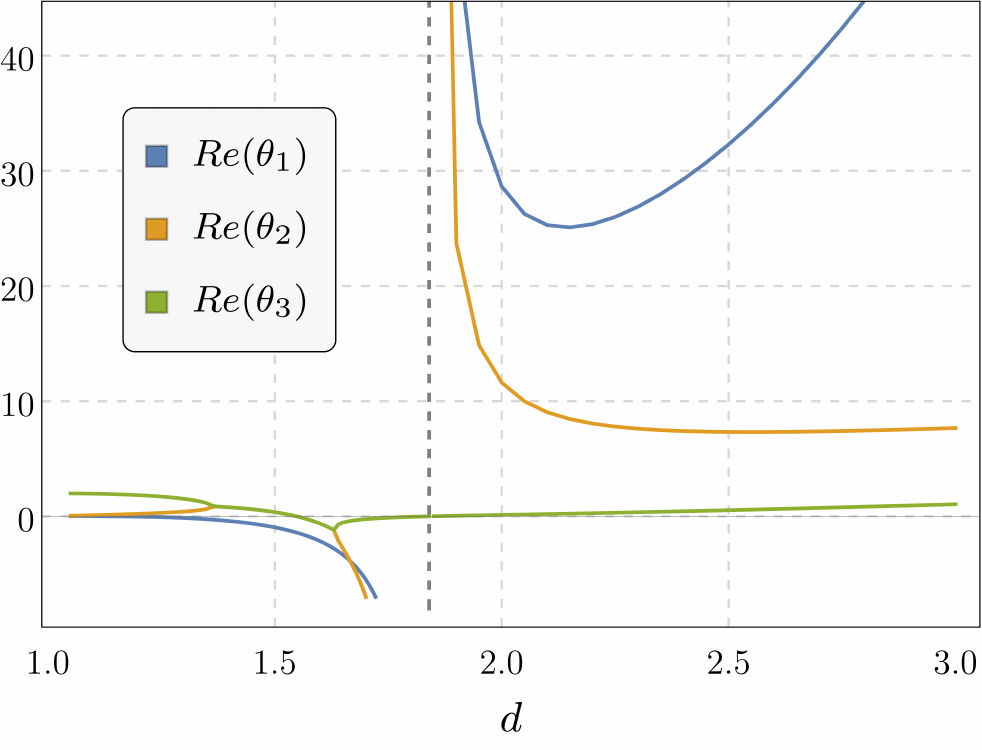}$\;\;$
\includegraphics[scale=0.42]{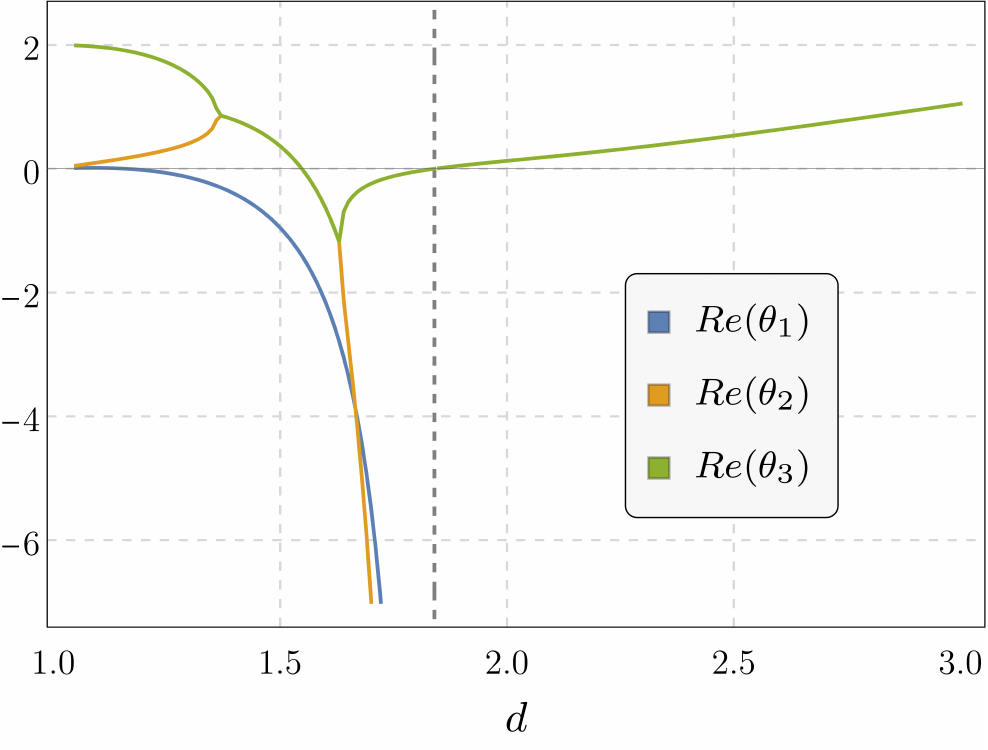}
\caption{Variation of the real part of the critical exponents with the spatial dimension $d$, for $N=2$. The figures show two different zooms. \label{Fig:truncation2bis}}
\end{figure}

The behaviour of the (real part of the) critical exponents as a function of $d$ for $N=2$ is shown in~Fig.~\ref{Fig:truncation2bis}. For $1<d\lesssim 1.55$, The NGFP emerging from the GFP is a saddle point with two relevant directions. For $ 1.55\lesssim d<d_{crit}$, the NGFP has three IR-irrelevant directions, \textit{i.e.} it is IR-attractive in the theory sub-space $\{g_k,f_0(k),f_2(k)\}$. Two of the critical exponents are complex conjugates for $1.37 \lesssim d \lesssim 1.63$ and real elsewhere. Finally, when $d$ exceeds the critical dimension $d_{crit}$, the NGFP at $g_*>0$ is replaced by the one at $g_*<0$, which has three IR-relevant directions and is therefore UV-attractive in the theory sub-space under consideration. 

\begin{figure}
\includegraphics[scale=0.35]{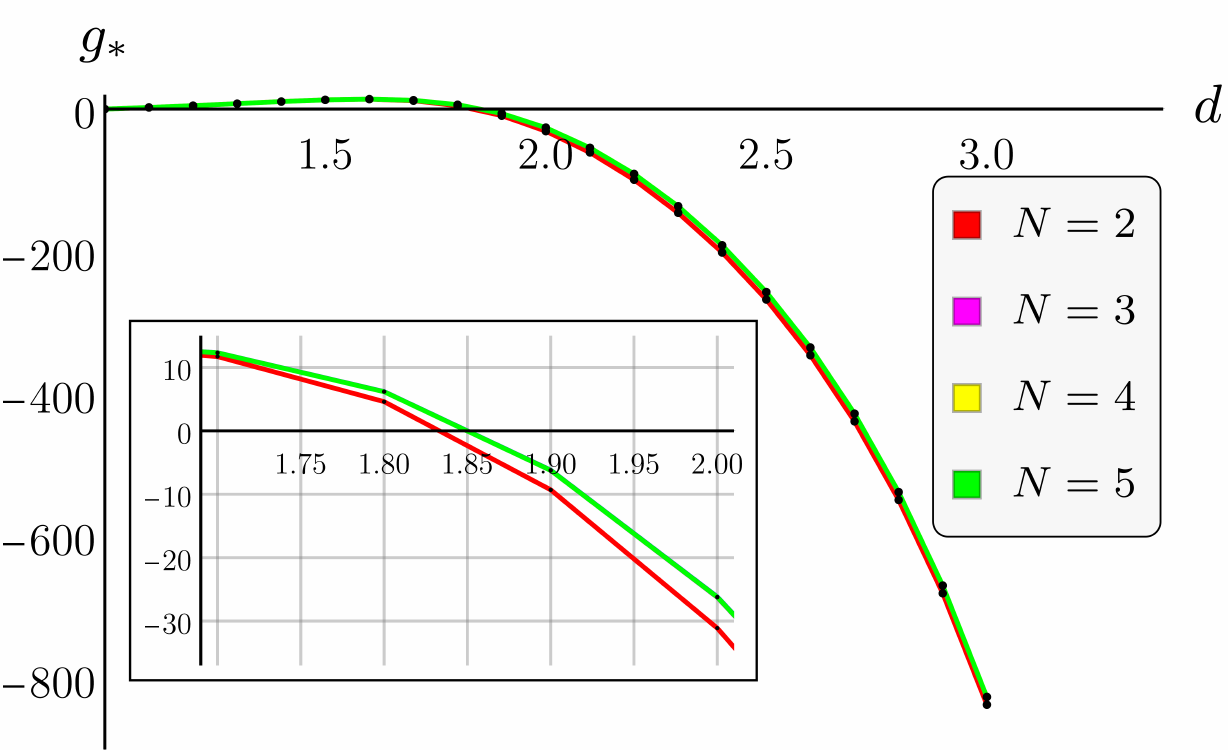}$\quad$
\includegraphics[scale=0.35]{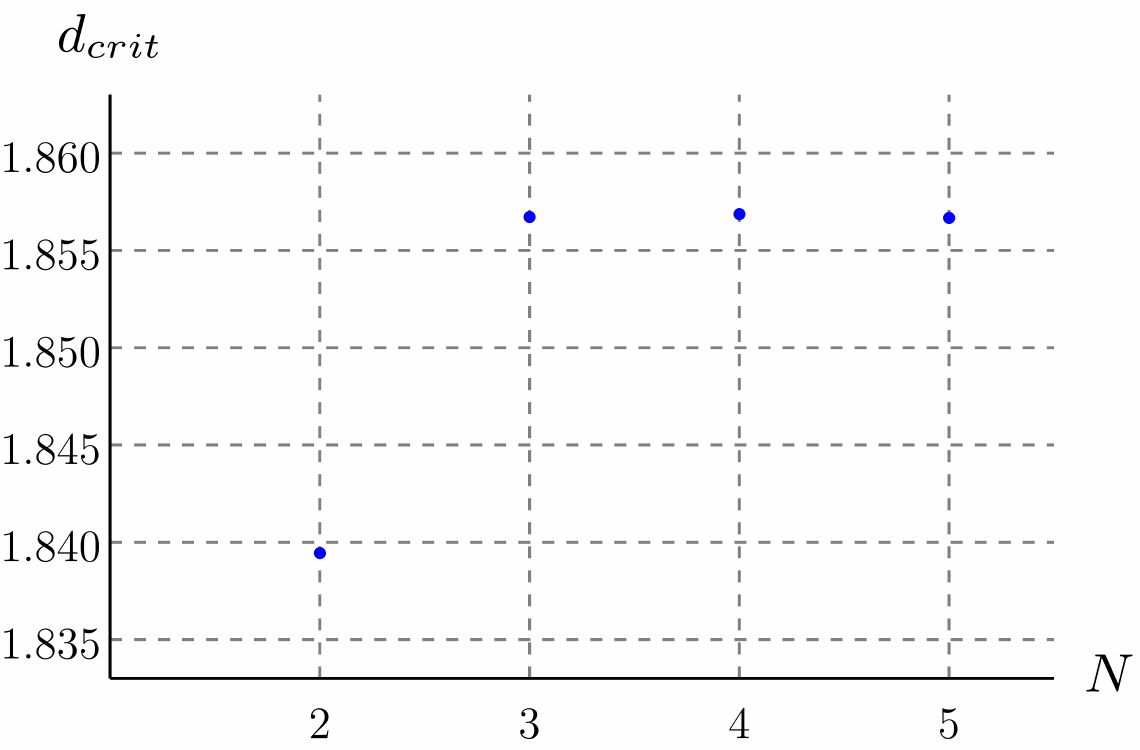}
\centering{}\caption{Position of the $g_{\ast}$-coordinate of the NGFP (left panel) and critical dimension $d_{crit}$ as functions of the truncation order~$N$. For $N>2$, the curves describing the variation of the $g_*$-coordinate of the NGFP/QFP overlap and are almost indistinguishable. Accordingly, for $N>2$ the value of the critical dimension $d_{crit}$ does not undergo significant changes. \label{Fig:truncation3}}
\end{figure}

The existence of a critical dimension and the related disappearance of the NGFP at $g_{*}>0$ could in principle be an artifact of the truncation. It is therefore necessary to investigate the fixed-point structure of the flow equations for larger truncation-orders~$N$. 
In higher-order truncations, up to the order $N=5$, the beta functions become very involved and highly singular. In particular, the denominators of all beta functions are proportional to the couplings $f_{i}$ with $i\geq2$, making them singular on the hypersurface $f_{i}=0$. 
The existence of fixed-point solutions and their connection with the NGFP found in $d=1+\epsilon$ can nevertheless be studied numerically. The extension of the NGFP at $g_*<0$ to larger theory sub-spaces, with $2<N\leq 5$, is a point with coordinates $f_{i,*}=0$ for all $f_{i}$ with $i\geq2$. In other words, increasing the truncation-order $N$, the NGFP at $g_*<0$ moves to an hypersurface where the beta functions $\beta_i$ are singular. 
Thus, for $N>2$ the ``fixed point'' at $g_\ast<0$ is no longer a solution to the equations $\beta_i=0$, rather it corresponds to a point in theory space where the beta functions are of the form $0/0$. 
Depending on how the couplings $f_i(k)$ with $i\geq 2$ approach zero, this point could still act as an attractor for specific RG trajectories. In other words, once the ansatz for the EAA is extended adding operators of the form $H^{2n}$ with $n>1$, the NGFP at $d>d_{crit}$ found for $N=2$ becomes a quasi-fixed-point (QFP). 
Additional fixed points at $g_*>0$ seem not to be generated. In particular, in analogy with the case $N=2$, there exists a critical dimension $d_{crit}\simeq 1.8$ where the $g_*$-coordinate of the NGFP found in $d=1+\epsilon$ vanishes (cf. Fig.~\ref{Fig:truncation3}). The value of the critical dimension $d_{crit}$ appears very stable: as shown in the right panel of Fig.~\ref{Fig:truncation3} and in Table~\ref{tab:critdim}, for $N>2$ and up to  $N=5$, the value of the critical dimension changes only to the fourth decimal digit.

\begin{table}
\begin{center}
\begin{tabular}{|c|c|c|c|c|}
\hline 
 & $N=2$ & $N=3$ & $N=4$ & $N=5$\tabularnewline
\hline 
$d_{crit}$ & 1.83944 & 1.85672 & 1.85687 & 1.85667\tabularnewline
\hline 
\end{tabular}
\caption{Value of the critical dimension as function of the truncation order $N$. \label{tab:critdim}}
\end{center}
\end{table}

\subsection{Considerations on the relation with string theory in the UV}\label{sec:commentonstringyuv}

Let us now comment on the relation between our effective actions, parametrized by the IR coefficients $c_n$, and string theory in the UV.

Different string models correspond to specific RG trajectories that determine the coefficients $c_n$ in the IR from a relevant deformation of a UV fixed point. At present, assessing whether an effective action corresponds to a given string theory entails comparing it to the result of a string computation. In principle, a set of coefficients $c_n$ could arise from both critical and non-critical strings, bosonic or heterotic~\cite{Hohm:2019jgu}. A priori, one would not expect that the map between Meissner-Hohm-Zwiebach fixed points and string theories be one-to-one. In particular, different string theories could yield the same IR coefficients $c_n$.

The consistency requirements of analyticity and the existence of a UV fixed point single out Eq.~\eqref{eq:string-frame_IR_EA} among our solutions. Thus, a comparison with known curvature corrections in a particular, critical string model would require resumming our $\epsilon$-expansion in order to estimate the coefficients in higher dimensions. Alternatively, one would need to find the general analytic solution of the flow equations compatible with a UV fixed point in a critical dimension. Both tasks are however highly involved, and go beyond the purpose of our work.

On the other hand, provided that such a solution exists, one can use it to investigate some phenomenological implications of T-duality, independently of the stringy details of the theory. This shall be the focus of the next subsection, where we study potential cosmological implications of T-duality in $(2+\epsilon)$-dimensions. See also~\cite{Basile:2021krk} for a study on the existence of de Sitter vacua in this setting.

\subsection{Cosmological implications}\label{sec:phys}

Let us now briefly comment on some potential implications of our findings. Specifically, we discuss quantum-corrected cosmological solutions to the field equations for the dilaton and the scale factor stemming from the quantum effective action in~Eq.~\eqref{eq:string-frame_IR_EA}.

To begin with, let us observe that T-duality implies that $\alpha'$-corrections only involve powers of first derivatives, and thus yield second-order equations of motion, avoiding Ostrogradsky-like instabilities. Moreover, let us recall that the IR parameter $G_{\text{N}}$ combines with the dilaton vacuum expectation value $\phi_0$ to build the observed Newton constant $G_{\text{obs}} \equiv g_s^2 \, G_{\text{N}}$, where the string coupling $g_s = e^{\phi_0} \ll 1$ in order for our approximations to be consistent. 

If $\Lambda < 0$ the IR Lagrangian features a branch cut in $H$ and is generally complex-valued, and thus unphysical. If $\Lambda > 0$ there are no de Sitter solutions to the field equations, since they would be incompatible with the Hamiltonian constraint. Specifically, as we have shown in~\cite{Basile:2021krk}, both string-frame and Einstein-frame de Sitter solutions are excluded for our effective actions~$\Gamma_\mathrm{string}$. This result appears to support the no-de Sitter Swampland conjecture~\cite{Obied:2018sgi, Ooguri:2018wrx, Garg:2018reu}. 

\begin{figure}
\hspace{-0.2cm}\includegraphics[scale=0.41]{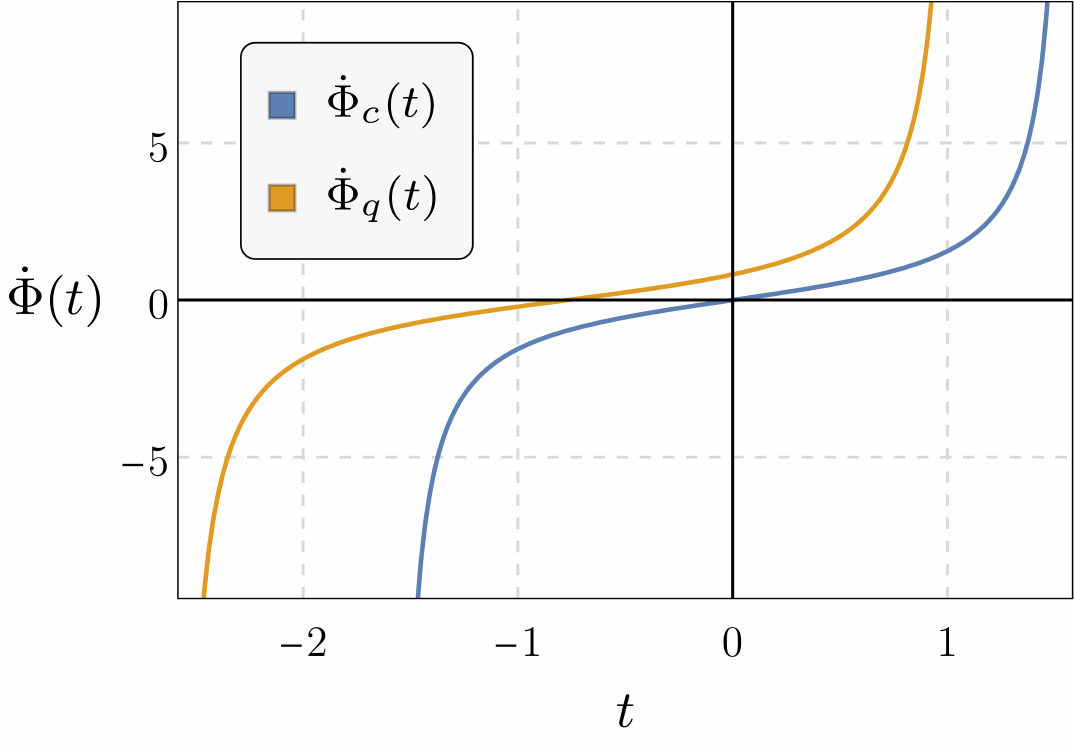}$\,\,$
\includegraphics[scale=0.41]{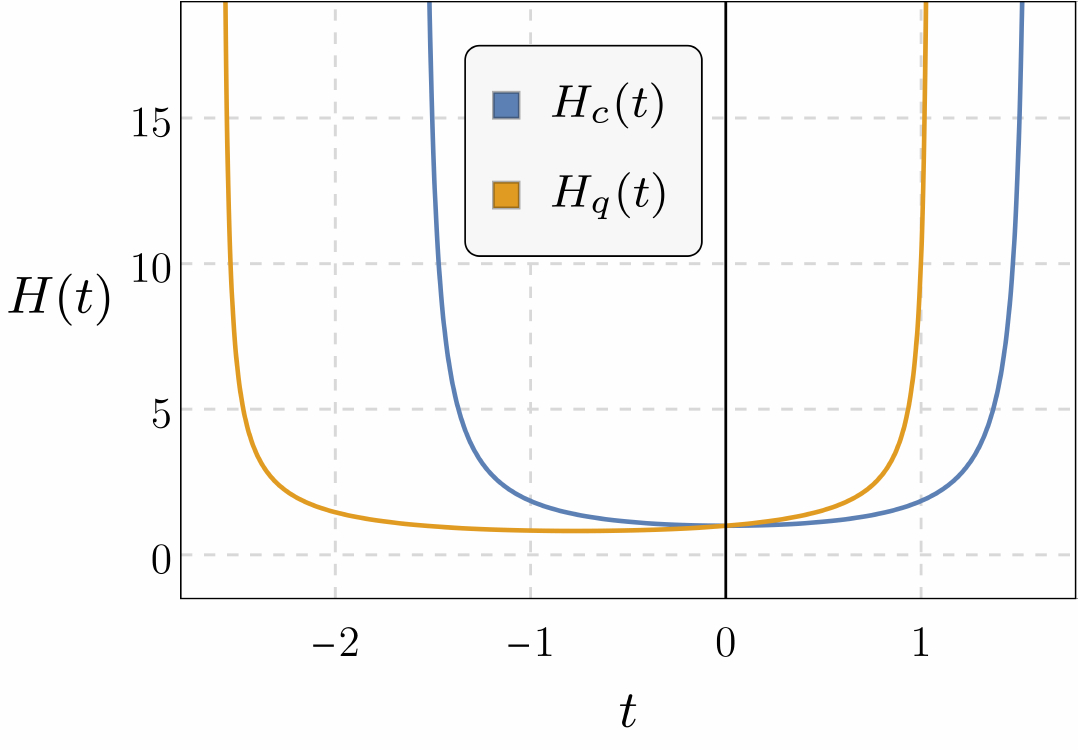}
\caption{Scaling of the first time derivatives of the dilaton $\Phi(t)$ (left panel) and of the Hubble parameter $H(t)$ (right panel). All quantities are in units of $\Lambda$. The ``classical'' (blue line) and quantum-corrected (orange line) curves share a qualitatively similar shape. The quantum-corrected curve, which arises from the effective action in Eq.~\eqref{eq:string-frame_IR_EA}, also appears to be shifted in time with respect to the classical one. \label{Fig:cosmostrings}}
\end{figure}

One can nonetheless consider more general cosmological solutions. However, solving the field equations stemming from the action in Eq.~\eqref{eq:string-frame_IR_EA} shows that the quantum-corrected solution is not qualitatively altered by the higher-derivative $\alpha'$-corrections encoded in the function $L(s)$, as depicted in Fig.~\ref{Fig:cosmostrings}. 

\section{Conclusions}\label{sec:conclusions}

In this paper we have investigated the application of FRG techniques within the framework of string theory. While the former appear to offer, at least in principle, quantitative tools to understand non-perturbative physics, in practice they offer only approximate pictures within truncations of the full theory space. On the other hand, the latter provides recipes to systematically derive curvature corrections, but the resulting computations are highly involved. Furthermore, genuinely stringy effects such as T-duality provide strong constraints on curvature corrections, to the extent that, in suitable time-dependent backgrounds, they are entirely classified by a single function. Therefore, it is conceivable that applying FRG methods to these settings could circumvent these technical issues, at least partially.

Motivated by this enticing prospect, in Section~\ref{sec:t-duality_frg} we have studied the flow of mini-superspace effective actions of the type derived in~\cite{Meissner:1991zj, Meissner:1996sa, Hohm:2015doa, Hohm:2019ccp, Hohm:2019jgu} via FRG methods, determining $\alpha'$-corrections to all-orders. In particular, we have obtained an asymptotically safe exact solution in general spacetime dimension $D$, which however features unphysical low-energy properties, and a phenomenologically viable solution within an $\epsilon$-expansion for $D = 2 + \epsilon$. However, within polynomial truncations of the theory space, the latter solution seems to disappear in higher dimensions. This could be due, among other possibilities, to a truncation artifact, a limitation of our mini-superspace approach, or to the importance of $g_s$-corrections, which we have neglected. Furthermore, we have investigated cosmological solutions stemming from the EAA found within the $\epsilon$-expansion, finding no qualitative deviation from the classical case defined in Eq.~\eqref{eq:classical_IR_terms}. In particular, the absence of de Sitter solutions garners further evidence for some Swampland conjectures~\cite{Obied:2018sgi, Ooguri:2018wrx, Garg:2018reu}.

All in all, determining curvature corrections in string theory is paramount to shed light on a number of aspects of high-energy physics, including singularities whose resolution would presumably involve all-order effects. While still lacking a complete non-perturbative formulation, the theory affords regimes in which some corrections can be systematically computed to arbitrarily high orders in a well-defined fashion. While earlier efforts in this respect were crucial in establishing string theory as a candidate for a quantum theory of gravity, a number of technical obstacles led a large portion of the community to focus on supersymmetric constructions, whose remarkable non-renormalization properties simplify matters dramatically. However, the phenomenological necessity to formulate models where supersymmetry is either broken or absent entails numerous subtleties, and the complete picture is still to be unveiled. Based on this state of affairs, we are therefore compelled to seek instructive lessons and methods beyond those provided by effective field theory, and to this end an approach based on stringy formulations and symmetries appears potentially fruitful. The results that we have presented in this paper constitute a first step in this direction, and we would like to explore these intriguing ideas further in future work.

\section*{Acknowledgements}

The authors would like to thank A. Sagnotti and F. Saueressig for discussions and comments on the manuscript. IB is grateful to F. Bascone and D. Bufalini  for sharing some useful references. During the development of this work, IB was supported in part by Scuola Normale Superiore and by INFN (IS CSN4-GSS-PI), while AP was supported by the Alexander von Humboldt Foundation. AP acknowledges support by Perimeter Institute for Theoretical Physics. Research at Perimeter Institute is supported in part by the Government of Canada through the Department of Innovation, Science and Economic Development Canada and by the Province of Ontario through the Ministry of Colleges and Universities. The work of I.B.\ was supported by the Fonds de la Recherche Scientifique - FNRS under Grants No.\ F.4503.20 (``HighSpinSymm'') and T.0022.19 (``Fundamental issues in extended gravitational theories''). AP is also grateful to Scuola Normale Superiore for hospitality during the early stages of development of this work.






\bibliographystyle{JHEP}

\bibliography{t-dual.bib}

\end{document}